\newif\ifnumlines
\numlinesfalse
\documentclass[preprint2]{aastex631}

\graphicspath{{./}{figures/}}

\usepackage{amsmath}
\usepackage{chemformula}
\let\tablenum\relax
\usepackage{siunitx}
\usepackage[verbose]{placeins}
\usepackage{graphicx}
\usepackage{chngpage}
\usepackage{xr}

\begin{document}

\title{Statistical geochemical constraints on present-day water outgassing as a source of secondary atmospheres on the TRAPPIST-1 exoplanets}

\author[0000-0003-2457-2890]{Trent B. Thomas}
\affiliation{Department of Earth and Space Sciences, University of Washington, Seattle, WA, USA}
\affiliation{Astrobiology Program, University of Washington, Seattle, WA, USA}
\affiliation{Virtual Planetary Laboratory, University of Washington, Seattle, WA, USA}

\author[0000-0002-1386-1710]{Victoria S. Meadows}
\affiliation{Astrobiology Program, University of Washington, Seattle, WA, USA}
\affiliation{Virtual Planetary Laboratory, University of Washington, Seattle, WA, USA}
\affiliation{Department of Astronomy, University of Washington, Seattle, WA, USA}

\author{Joshua Krissansen-Totton}
\affiliation{Department of Earth and Space Sciences, University of Washington, Seattle, WA, USA}
\affiliation{Astrobiology Program, University of Washington, Seattle, WA, USA}
\affiliation{Virtual Planetary Laboratory, University of Washington, Seattle, WA, USA}

\author[0000-0002-2587-0841]{Megan T. Gialluca}
\affiliation{Astrobiology Program, University of Washington, Seattle, WA, USA}
\affiliation{Virtual Planetary Laboratory, University of Washington, Seattle, WA, USA}
\affiliation{Department of Astronomy, University of Washington, Seattle, WA, USA}

\author[0000-0002-0413-3308]{Nicholas F. Wogan}
\affiliation{Virtual Planetary Laboratory, University of Washington, Seattle, WA, USA}
\affiliation{NASA Ames Research Center, Moffett Field, CA 94035, USA}

\author[0000-0001-5646-120X]{David C. Catling}
\affiliation{Department of Earth and Space Sciences, University of Washington, Seattle, WA, USA}
\affiliation{Astrobiology Program, University of Washington, Seattle, WA, USA}
\affiliation{Virtual Planetary Laboratory, University of Washington, Seattle, WA, USA}

\begin{abstract}
The TRAPPIST-1 planetary system is observationally favorable for studying if planets orbiting M stars can retain atmospheres and host habitable conditions. Recent JWST secondary eclipse observations of TRAPPIST-1 c rule out a thick \ch{CO2} atmosphere but do not rule out atmospheric water vapor or its photochemical product, oxygen. Given the high expected escape rate, maintenance of atmospheric water vapor would require a present-day water source, such as volcanic outgassing. Here, we simulate water outgassing on the TRAPPIST-1 planets over a broad phase space based on solar system terrestrial bodies. We then apply two filters based on observation and geochemistry that narrow this phase space and constrain the plausible outgassing scenarios. For all seven TRAPPIST-1 planets, we find that the water outgassing rate is most likely $\sim$0.03x Earth's but has upper limits of $\sim$8x Earth's. The allowed range also implies low, Mars-like magma emplacement rates and relatively dry, Earth-like mantles, although mantle water mass fractions up to 1 wt\% are possible. We also present scenarios with magma emplacement rates similar to Mars, Earth, and Io, resulting in different preferred mantle water content and outgassing rates. We find that water outgassing rates are potentially high enough to balance water escape rates, providing a theoretical pathway for the TRAPPIST-1 planets to maintain surface water or water-vapor-containing atmospheres over long timescales. The bounds on outgassing rates and interior properties can be used in atmospheric chemistry and escape models to contextualize future observations of the TRAPPIST-1 planets, and may be applicable to other terrestrial exoplanets.
\end{abstract}

\section{Introduction}

The TRAPPIST-1 planetary system is a natural laboratory for studying habitability. This ultracool M dwarf star harbors seven transiting, terrestrial-sized  planets, which span orbital distances interior to, within, and beyond the star's habitable zone \citep{Gillon.2017}. Most of the planets are remarkably similar to Earth in mass and radius but have a consistently lower density than the terrestrial planets in our Solar System \citep{Agol.2021}. Initial observations with the Hubble Space Telescope (HST) ruled out H$_2$-dominated primary atmospheres for the first six planets from the star \citep{Wit.2016,Wit.2018}. M dwarf stars are $\sim3/4$ of the stars in the galaxy \citep{Bochanski.2010}, so the potential habitability of the TRAPPIST-1 planets may be indicative of potentially habitable worlds elsewhere. To that end, it is critical to understand whether planets orbiting ultracool dwarf stars can sustain secondary, outgassed atmospheres and host habitable surface conditions \citep[e.g.][]{Shields.201604g, Meadows.2018mts, Krissansen-Totton.2024}. We now have the opportunity to investigate these possibilities by observing and characterizing the TRAPPIST-1 planets.

Recent JWST observations have placed constraints on the presence and nature of atmospheres on the TRAPPIST-1 planets, with no atmosphere or tenuous atmospheres being the most favored explanations, although some denser atmospheres may still be consistent with the data. Secondary eclipse observations have been made for the two innermost planets, TRAPPIST-1 b and c \citep{Greene.2023, Zieba.2023}. For TRAPPIST-1 b, \citet{Greene.2023} conclude that the 15 $\mu$m secondary eclipse measurement is most consistent with an airless rock. In contrast, clear-sky and cloudy Venus-like atmospheres, and \ch{O2}-dominated atmospheres with more than 0.5 bar \ch{CO2}, were ruled out at over 6-$\sigma$ \citep{Greene.2023}, and atmospheres greater than 0.3 bar with at least 100 ppm \ch{CO2} were ruled out at 3-$\sigma$ \citep{Ih.2023}. However, a dense atmosphere may not be completely ruled out, as a preliminary combined analysis of JWST 15$\mu$m and 12.8$\mu$m secondary eclipse observations appears to be consistent with models of both an airless, fresh ultramafic surface -- possibly indicative of active volcanism -- or possibly a dense \ch{CO2} atmosphere with a photochemical, haze-induced temperature inversion \citep{Ducrot.2023}.   

For TRAPPIST-1 c, the  \citet{Zieba.2023} 15 $\mu$m secondary eclipse measurement is also consistent with a range of airless and atmospheric scenarios. They found that the best fits to the data were for models with an airless rock or a thin (e.g., 0.1 bar) \ch{O2}-\ch{CO2} atmosphere. They also found that Venus-like atmospheres of $\geq 10$ bar, and \ch{O2}-\ch{CO2} atmospheres of greater than 1 bar surface pressure with more than 1000 ppm \ch{CO2}, were ruled out at greater than 2.6-$\sigma$. \citet{Lincowski.2023} expanded the compositional range of model atmospheres, finding that atmospheres that contain water vapor are also consistent with the data to within 1.7-1.8-$\sigma$, including a 0.1 bar \ch{O2}, 100 ppm \ch{CO2}, 10\% water vapor atmosphere, or a $\leq 3$ bar steam atmosphere. Transmission spectroscopy of TRAPPIST-1 b, which would be complementary to the secondary eclipse observations, has yet to yield the sensitivity needed for molecular detection in a secondary atmosphere, due to the challenges of removing stellar contamination from the spectra \citep{Lim.2023}. For the rest of the TRAPPIST-1 planets, which are likely too cool to have observable mid-infrared (MIR) secondary eclipses, initial JWST transmission spectroscopy has been obtained and these data are still under analysis \citep{JWST2017:1201GTO,JWST2021:1981GO,JWST2021:2589GO,JWST2022:2759GTO}. So, whether or not these planets have secondary atmospheres is still unconstrained.

If the putative present-day atmospheric water on the TRAPPIST-1 planets is the result of delivery during planet formation and subsequent loss over time, then a large initial water endowment is required for any to remain today \citep[e.g., over 50 Earth oceans would be required on the 3 inner planets;][]{Gialluca.2024}. The large initial water endowment is required because water in the atmosphere would have been subject to intense atmospheric escape driven by the high luminosity and radiation from the host star during the lifetime of the system. Specific processes that would drive ocean and atmospheric escape include the long-lived super-luminous pre-main-sequence phase \citep{Luger.2015, Bolmont.2016, Gialluca.2024}, the high baseline stellar XUV flux and stellar wind during the main sequence \citep{Bourrier.2017, Dong.2018, Gialluca.2024}, and frequent stellar flares throughout the lifetime of the star \citep[][]{Amaral.2022}. If even traces of water vapor are confirmed in the atmosphere of TRAPPIST-1 c, it is important to know where this water came from and how it could have survived this intense atmospheric escape.

Alternatively, instead of originating from a large initial endowment, any current water in the atmospheres of the TRAPPIST-1 planets could be supplied by continuous outgassing from the planets' interiors. This is especially true for simulations of more tenuous 0.1 bar O$_2$ atmospheres with greenhouse gases such as water vapor, that produce better fits to the observations of the two innermost planets \citep[][Gillon et al., forthcoming]{Ih.2023,Lincowski.2023}. These tenuous atmospheres are likely to have short (less than a million year) lifetimes against atmospheric loss \citep{Dong.2018}, and so must be continuously replenished to persist to present day. These atmospheres also provide poor day-night heat transport, potentially resulting in freezing night side temperatures that condense and trap water vapor \citep{Lincowski.2023}.   

Consequently, outgassed water on the TRAPPIST-1 planets would have important implications for their atmospheric evolution and habitability. On Earth, water vapor is the primary constituent of volcanic gases \citep[][p. 203]{Catling.2017}, so it is reasonable to explore how water could be outgassed on the TRAPPIST-1 planets, given that their densities are consistent with terrestrial composition \citep{Agol.2021}. If water is stored in the planets' interiors before being outgassed to the atmosphere, then the outgassing bottleneck decreases the availability of water vapor for atmospheric escape during the lifetime of the system. This process would help to preserve the planetary water reservoir and extend the availability of water vapor in the planetary atmosphere \citep[e.g.,][]{Moore.2020}. Moreover, for the habitable zone planets, if the water outgassing rate is high enough, it could balance atmospheric escape \citep{Dong.2018, Gialluca.2024} and create steady state surface water that is potentially long-lived. 

Given that the TRAPPIST-1 planets have densities typically 5-10\% lower than similarly-sized solar system terrestrial planets \citep{Agol.2021}, higher water outgassing rates than Earth's might be expected. The low densities could be partially explained by a high interior water content, which may contribute to water-rich magmas and enhance water outgassing. Ultimately, if atmospheric water exists on the TRAPPIST-1 planets, it must come from either 1) an initial water endowment that was lost over time, or 2) continuous replenishment from a source like volcanic outgassing, or a combination of both.

The goal of this paper is to explore the likelihood of water outgassing as a potential source for present-day water on the TRAPPIST-1 planets by addressing the following question: what are plausible bounds on the present-day water outgassing rate? To address this question, we develop a modeling approach to statistically search a broad parameter space of possible present-day scenarios on the TRAPPIST-1 planets given their unconstrained geological properties. This model can help in the interpretation of atmospheric observations of the TRAPPIST-1 planets by identifying which scenarios are not only consistent with the observational data, but are also consistent with geochemical constraints based on solar system knowledge. If the TRAPPIST-1 planets are found to have no significant exterior water, the model also sets constraints on outgassing that may be applicable to observations of other planetary systems in the future. This approach is complementary to planetary evolution models, which yield valuable insights but may be less suited to exploring the broad phase space of present-day possibilities.

\section{Methods}

To estimate plausible outgassing rates on the seven TRAPPIST-1 planets, we first develop a theoretical outgassing model that explores a large parameter space, then we filter these results to identify the most plausible outgassing ranges by considering observational, empirical, and theoretical constraints on the TRAPPIST-1 planets.

\subsection{Theoretical model}

\begin{figure*}[htbp]
    \centering
    \includegraphics[width=0.8\textwidth]{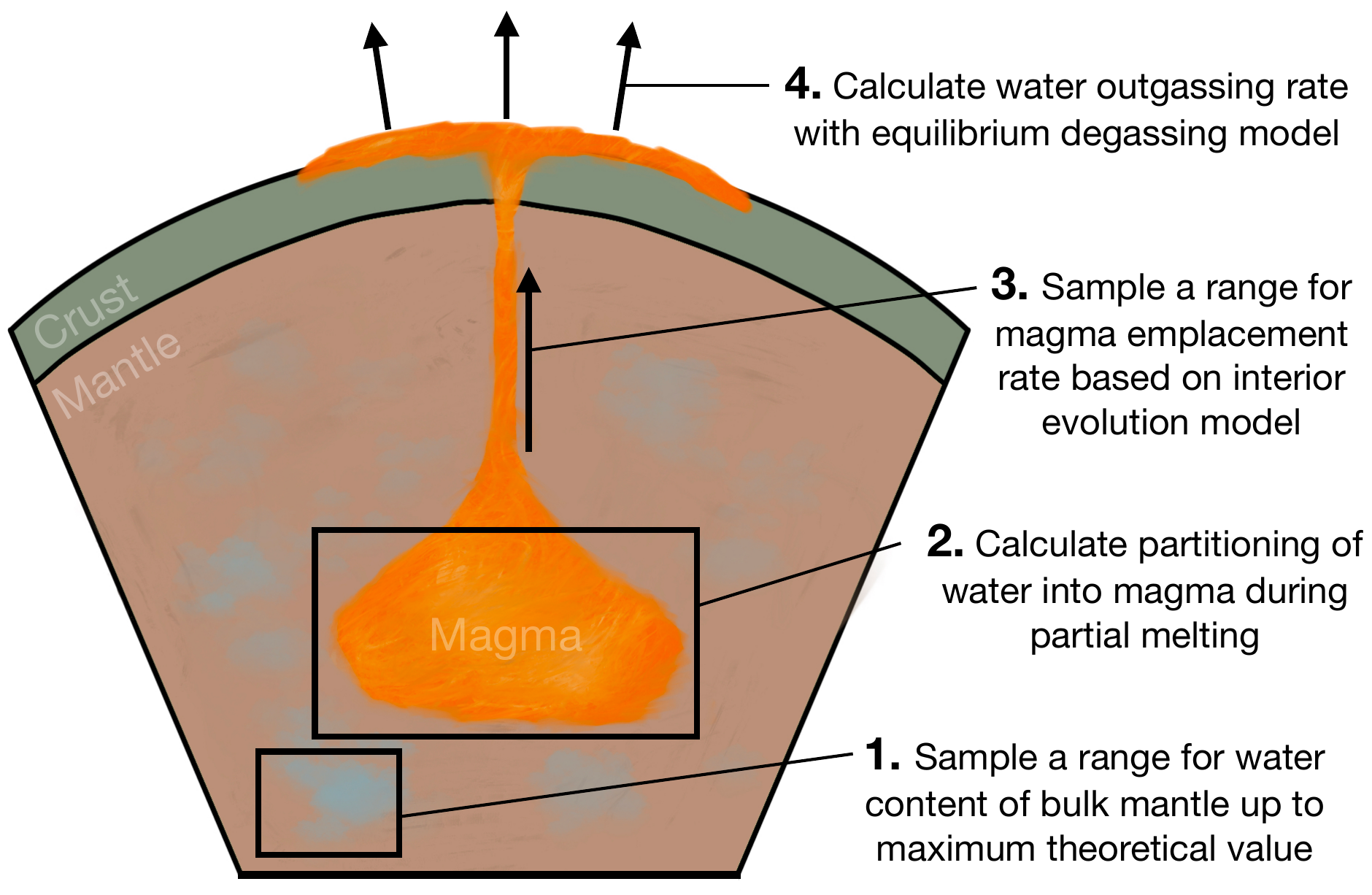}
    \caption{Schematic diagram of steps in the theoretical model. Further description of key parameters, sampled ranges, and calculation procedures is found in Methods.}
    \label{schematic}
\end{figure*}

In this section, we describe the theoretical model used to estimate modern rates of water outgassing on the TRAPPIST-1 planets, summarized in Figure \ref{schematic}. Our model is simpler in comparison to fully-coupled atmosphere-interior evolution models \citep[e.g.][]{Krissansen-Totton.2022}, but this design choice allows us to efficiently explore a large parameter space of possible present-day scenarios and generate statistical inferences. First, we estimate how much water could reasonably be stored in the interior of the TRAPPIST-1 planets based on theoretical studies that are informed by the mineralogy of solar system planets. Second, we describe an interior melting model that we use to determine how much water could be dissolved in interior magmas. Third, we estimate the rate at which magma is generated in the interior and delivered to the surface (i.e. the magma emplacement rate). Fourth, we estimate water outgassing rates on the TRAPPIST-1 planets using an equilibrium magmatic outgassing model that calculates the partitioning of gases (including water vapor) between the atmosphere and surface magma.

\subsubsection{Interior water content}\label{sec:interior_water}

In our model, we consider mantle-derived magmas as the ultimate source for outgassed water on the TRAPPIST-1 planets. Once mantle water is sourced to the exterior, it may be recycled between the surface and crust, but this does not provide additional water and thus does not help replenish water lost to space on long timescales. Therefore, we must know the mantle water mass fraction, $m_{\ch{H2O}}^{mantle}$, to begin our calculation. The value of $m_{\ch{H2O}}^{mantle}$ for the TRAPPIST-1 planets is unknown, and indirect constraints on this value, via e.g. density, are subject to degeneracies in the internal structure and composition \citep{Agol.2021}, which will be discussed later. For these reasons, we rely on theoretical studies, informed by Earth's properties, to estimate a plausible range of $m_{\ch{H2O}}^{mantle}$, including maxima for mantle water storage. 

To construct models of the TRAPPIST-1 planets, which have densities comparable to, but slightly lower than Earth's, we first consider the amount and mode of water storage in Earth's mantle. Earth's mantle contains $m_{\ch{H2O}}^{mantle} = 0.049-0.24$ wt\% water \citep{Ohtani.2020}, which is stored in a combination of hydrous minerals and nominally-anhydrous minerals (NAMs) \citep[e.g., reviewed by][]{Ohtani.2015, Ohtani.2020}. Hydrous minerals can store over 10 wt\% water by incorporating \ch{OH} or \ch{H2O} directly in their crystal structure \citep[see Tables 1 and 2 in][and references within]{Ohtani.2015}. Examples include serpentine, brucite, and numerous high pressure ``alphabet phase'' \ch{Mg}-silicate minerals. On the other hand, NAMs  rarely exceed 1 wt\% water, and store water by incorporating \ch{OH} groups as impurities in the gaps of their crystal structure. Examples include olivine, wadsleyite, ringwoodite, and perovskite. Despite the superior water storage capacity of hydrous minerals, Earth's deep mantle water is mostly stored in NAMs because hydrous minerals are typically unstable at the high temperatures and pressures found deep in the interior \citep{Bolfan-Casanova.2005}. However, differences in size, composition, and interior structure could allow other planets to have different proportions of these water-bearing minerals, and thus higher or lower $m_{\ch{H2O}}^{mantle}$ than Earth.

If we consider the multi-layered terrestrial planetary interior model of \citet{Shah.2021}, we can constrain the maximum possible $m_{\ch{H2O}}^{mantle}$ for the TRAPPIST-1 planets to $\sim3$ wt\%. \citet{Shah.2021} account for mantle hydration via both hydrous minerals and NAMs, including olivine polymorphs, brucite, perovskite, and post-perovskite, using available experimental data or molecular dynamics simulations when available. They find that the maximum capacity for total internal water storage on terrestrial planets is 6 wt\%. However, most of this water is predicted to be stored within iron hydride in the core where it is unlikely to contribute to surface degassing. \citet{Shah.2021} find that $m_{\ch{H2O}}^{mantle}$ is maximized when planets have high interior magnesium concentration relative to iron. For the masses of the TRAPPIST-1 planets, assuming they are magnesium-rich, the upper limit on mantle water mass as a fraction of total planet mass is $\sim2$ wt\% \citep[][fig. 7, top middle panel]{Shah.2021}. Scaling this according to the mass of the mantle, which is as low as 65\% of the total planet mass for the corresponding magnesium abundance \citep[][fig. 2]{Shah.2021}, results in an upper limit $m_{\ch{H2O}}^{mantle}$ of $\sim3$ wt\%.

Alternatively, \citet{Guimond.2023} consider water storage only in nominally anhydrous minerals, and estimate a more conservative upper limit on interior water storage of 1 wt\%. Whereas \citet{Shah.2021} assume a fixed mineralogy within idealized terrestrial planet mantles, \citet{Guimond.2023} specify bulk oxide content (e.g., \ch{MgO}, \ch{SiO2}, and \ch{FeO}) and use a thermodynamic model to calculate how these oxides speciate into different mineral phases. Importantly, \citet{Guimond.2023} only consider water storage in NAMs. They do not include hydrous mineral phases or storage in the core. \citet{Guimond.2023} find terrestrial planets should not exceed a total water mass fraction (including core mass) of 0.2 wt\% from mantle water storage alone, regardless of planet mass or compositional parameters (see their figure 5). Converting the total water mass fraction into $m_{\ch{H2O}}^{mantle}$ assuming a large core mass fraction of 80 wt\% \citep[i.e., larger than Mercury's 65 wt\%, which is the largest in the solar system;][]{Charlier.2019} yields the maximum $m_{\ch{H2O}}^{mantle}$ of 1 wt\%.

The interior water mass fractions of the TRAPPIST-1 planets are unlikely to be constrained through mass and radius observations \citep[even though both have been precisely measured to within $\sim$5\% accuracy;][]{Agol.2021}, highlighting the need for theoretical constraints. A surface water or ice layer has a more negative contribution to planet bulk density than the equivalent mass of H and O dissolved in interior silicates \citep[e.g.][]{Dorn.2021, Luo.2024}, which prevents us from extrapolating estimates of bulk water mass fraction from one reservoir to another. Furthermore, the size of each planetary water reservoir (e.g., core, mantle, exterior) is unknown. For example, it is possible that these planets do not have cores at all, which would immediately explain the low observed densities without the need for any additional low-density water \citep{Agol.2021}. Taken together, the above degeneracies and uncertainties regarding the TRAPPIST-1 planets' observed densities are precisely why theoretical constraints based on known geochemistry are required.

Given that the observations do not strongly constrain the $m_{\ch{H2O}}^{mantle}$ values, we use the above theoretical studies to generate a plausible range for $m_{\ch{H2O}}^{mantle}$ on the TRAPPIST-1 planets. We assume that the maximum value of $m_{\ch{H2O}}^{mantle}$ is 1 wt\% in the nominal model configuration, but we also test maximum values of 0.5 wt\% and 3 wt\% to assess model sensitivity. We assume that the minimum $m_{\ch{H2O}}^{mantle}$ is 0.01 wt\% to account for a scenario in which the TRAPPIST-1 planets are relatively dry. These values bracket the Earth's $m_{\ch{H2O}}^{mantle} = 0.049-0.24$ wt\% water \citep{Ohtani.2020}. In summary, we initially explore the broad range $m_{\ch{H2O}}^{mantle} = 0.01-3$ wt\% for the TRAPPIST-1 planets.

\subsubsection{Magma water content}

\begin{figure*}[htbp]
    \centering
    \includegraphics[width=\textwidth]{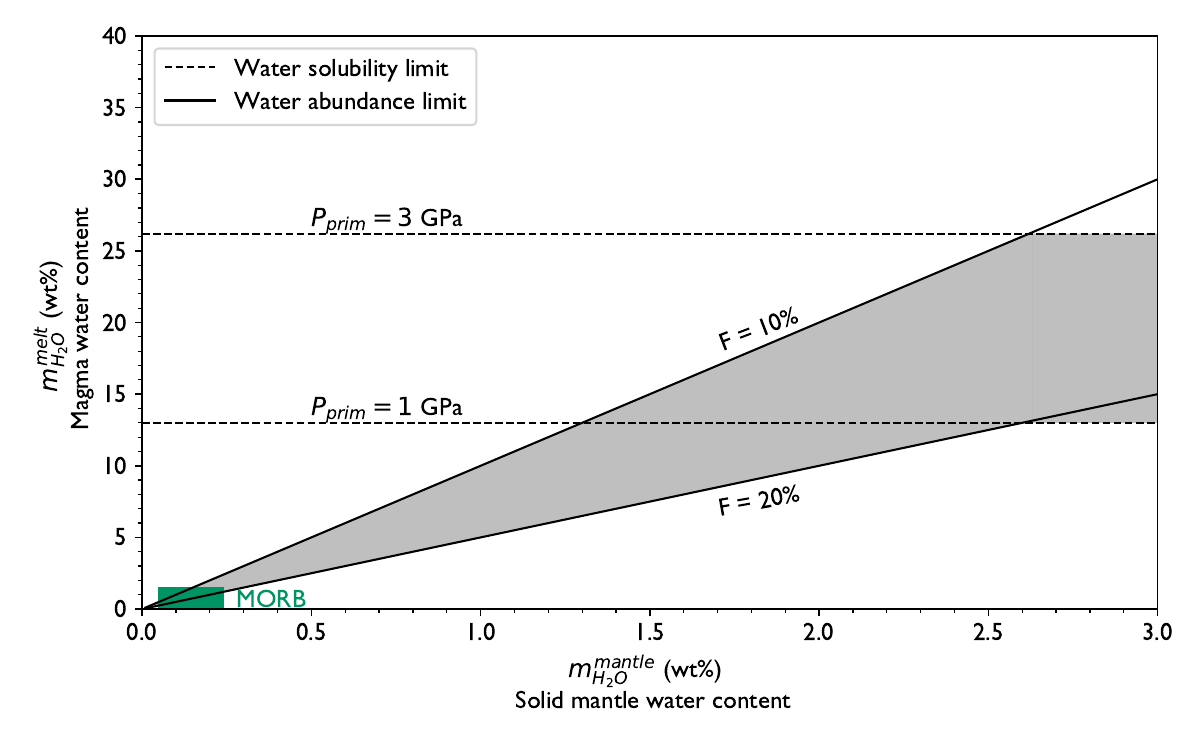}
    \caption{Water concentration in melt ($m_{\ch{H2O}}^{melt}$) as a function of bulk mantle water content ($m_{\ch{H2O}}^{mantle}$) considering water saturation and fractional melting. The dashed lines are the solubility-limited \ch{H2O} content of the magma under the maximum and minimum values of primary magma pressure, $P_{prim} = 1$ and 3 GPa. The black solid lines are the abundance-limited \ch{H2O} content of the magma assuming fractional melting under the maximum and minimum values of melt fraction, $F = 10$ and 20\%. For a given $m_{\ch{H2O}}^{mantle}$, $F$, and $P_{prim}$, the true value of $m_{\ch{H2O}}^{melt}$ is the minimum from water saturation or fractional melting. The gray shaded region is the range of possible values allowed based on this procedure. The green shaded region labeled "MORB" shows the range for mid-ocean ridge basalt magma water content \citep[$0.1 - 1.5$ wt\%;][Ch. 7]{Sigurdsson.2015} and Earth's mantle water concentration \citep[$0.049-0.24$ wt\%;][]{Ohtani.2020}.}
    \label{magma_main} 
\end{figure*}

\begin{deluxetable*}{lcl}
    \tablecaption{Outgassing model parameters \label{og_params}}
    \tablewidth{0pt}
    \tablehead{\nocolhead{} & \colhead{Nominal Range} & \colhead{Notes and References}}
    \tabletypesize{\footnotesize}
    \startdata
    Mantle water mass fraction, $m_{\ch{H2O}}^{mantle}$ & 0.01-1 wt\% & See main text \\
    Melt fraction, $F$ & 10--20\% & Based on MORBs $^{b}$ \\
    Pressure of primary magma formation, $P_{prim}$ & 1--3 GPa & Based on MORBs $^{b}$ \\
    Temperature of magma, $T$ & 873--1973 K & Based on Earth magmas $^{c}$ \\
    Surface pressure, $P_{surf}$ & 0.001--100 bar $^{a}$ & Broad TRAPPIST-1 planet range \\
    \ch{CO2} content in magma, $m_{CO_2}^{melt}$ & $10^{-5}$--$10^{-2}$ $^{a}$ & Based on Earth magmas $^{d}$ \\
    Oxygen fugacity of magma, $f_{O_2}$ & FMQ-4 to FMQ+5 $^{a}$ & Based on solar system observations $^{e}$\\
    \enddata
    \tablecomments{\\ $^{a}$ Denotes this parameter was sampled uniformly in log space. All others were sampled uniformly in linear space.\\ $^{b}$ See section on magma water content for further description. MORB = mid-ocean ridge basalt.\\ The following references and corresponding parameter ranges were compiled in \citet{Wogan.2020}:\\ $^{c}$ Coldest rhyolite magma to hottest komatiite magma \citep{Huppert.1984}.\\ $^{d}$ Approximate mass fraction range in Earth magma \citep{Wallace.2005, Wallace.2015, Anderson.2017, Voyer.2019}.\\ $^{e}$ Most reducing Martian meteorite to most oxidized Earth magma \citep[][p. 363]{Stamper.2014, Catling.2017}.}
\end{deluxetable*}

Assuming the above range of estimates for $m_{\ch{H2O}}^{mantle}$, we now calculate the corresponding water content of the interior magmas for the TRAPPIST-1 planets by considering the diversity of Earth's mantle-derived magmas and the processes that produce them. Magmas can have higher water content than the mantle minerals from which they are derived, as water preferentially partitions into magma during partial melting \citep{Ni.2016}. Earth's magmas that are produced by subduction processes, giving rise to island arc basalts or continental margin basalts, typically have the highest water fractions due to the recycling of liquid water from the surface \citep[][Ch. 7]{Sigurdsson.2015}. For the TRAPPIST-1 planets, we base our calculations on the mantle-derived magmas because they are the net source of water to the surface and atmosphere, whereas recycling processes (if present) are only responsible for the partitioning of water once it is at the surface or atmosphere. Therefore, we focus on the melting processes that generate Earth's mantle-derived mid-ocean ridge basalts (MORBs) and ocean island basalts (OIBs). These magmas typically have \ch{H2O} concentration of 0.1 - 1.5 wt\% due to preferential water incorporation in the melt \citep[][Ch. 7]{Sigurdsson.2015}, up to $\sim30$x Earth's bulk mantle water content \citep{Ohtani.2020}.

The water content in the melt, $m_{\ch{H2O}}^{melt}$, can be limited by either the amount of water in the melting rocks (water abundance limit) or by the pressure-dependent ability of water to dissolve in the magma (water solubility limit). The value of $m_{\ch{H2O}}^{melt}$ is then given by the minimum value calculated in these two limits. In the water abundance limit, we use the fractional melt model of \citet{Grott.2011}, which is based on known magma chemistry on Earth, and which \citet{Grott.2011} applied to Mars. Following their equation 6, $m_{\ch{H2O}}^{melt}$ as a function of $m_{\ch{H2O}}^{mantle}$ is given by 
\begin{equation}
m_{\ch{H2O}}^{melt} = \frac{m_{\ch{H2O}}^{mantle}}{F} \Bigl[1-(1-F)^{1/D}\Bigr]
\label{melt}
\end{equation}
where $F$ is the percentage of the mantle rocks that melt (i.e., local melt fraction), typically assumed to be 10-20\% for MORBs \citep[e.g.,][]{Bolfan-Casanova.2005}, and $D$ is the partitioning coefficient for water. $D$ is defined as the concentration of water remaining in the rocks divided by the concentration of water that dissolves in magma, so lower $D$ means more water in the magma. We assume $D = 0.01$ \citep{Katz.2003}, but it can be as low as 0.001 in some cases \citep{Aubaud.2004}; however, for either value of $D$, the term $(1-F)^{1/D}$ in Eq. \ref{melt} is $<<1$ for the range of local melt fractions we assume. So, $m_{\ch{H2O}}^{melt}$ is primarily a function of the first term in Eq. \ref{melt}, $m_{\ch{H2O}}^{mantle}/F$. We assume that the water concentration is homogeneous within the mantle water reservoir.

In the water solubility limit we use a model that is similarly based on known magma chemistry on Earth and was previously applied to Mars \citep{Katz.2003, Grott.2011}. The pressure-dependent saturation concentration of \ch{H2O} in the melt is given by
\begin{equation}
    m_{\ch{H2O}}^{melt} = \chi_1 P_{prim}^{\xi} + \chi_2 P_{prim}
    \label{saturation}
\end{equation}
where $\chi_1 = 12$ wt\% GPa$^{-\xi}$, $\chi_2 = 1$ wt\% GPa$^{-1}$, and $\xi = 0.6$ are empiricial constants. $P_{prim}$ is the pressure at which the primary magma is produced in the planet's interior, typically 1-3 GPa for MORBs on Earth \citep[][Ch. 1]{Sigurdsson.2015}. 

The water concentration in the melt is then taken as the minimum concentration calculated from the water abundance limit (Equation \ref{melt}) or water saturation limit (Equation \ref{saturation}). This relationship is shown in Figure \ref{magma_main} and the parameter ranges we explore are shown in Table \ref{og_params}. 

\subsubsection{Magma emplacement rate} \label{Q_methods}

\begin{deluxetable*}{llllllll}
    \tablecaption{TRAPPIST-1 model planet parameters \label{t1_params}}
    \tablewidth{0pt}
    \tabletypesize{\footnotesize}
    \tablehead{\nocolhead{} & \colhead{b} & \colhead{c} & \colhead{d} & \colhead{e} & \colhead{f} & \colhead{g} & \colhead{h}}
    \startdata
    Max. magma emplacement rate, $Q$ [km$^{3}$ yr$^{-1}$] & 8500 & 5840 & 1230 & 2159 & 2130 & 2344 & 1230 $^{a}$ \\
    Received XUV flux, $F_{\rm XUV}$ [erg s$^{-1}$ cm$^{-2}$] & 2935.13 & 1564.87 & 788.455 & 456.452 & 263.598 & 178.116 & 102.035 \\
    Energy-limited \ch{H} escape rate, $\Phi_{\ch{H}}$ [$10^{12}$ kg yr$^{-1}$] & 57.3 & 30.5 & 19.2 & 9.90 & 5.58 & 3.74 & 2.60 \\
    \enddata
    \tablecomments{$Q$ values are from \citet{Krissansen-Totton.2022}. $F_{\rm XUV}$ values are from \citet{Becker.2020}. See main text for description of $\Phi_{\ch{H}}$. \\ $^{a}$ No estimate is provided for TRAPPIST-1 h in \citet{Krissansen-Totton.2022}, so the upper limit is assumed to be the same as TRAPPIST-1 d due to their similar masses.}
\end{deluxetable*}

We now calculate the volumetric magma emplacement rate, $Q$ (km$^{3}$ yr$^{-1}$), on the TRAPPIST-1 planets. Magma emplacement, the volcanic deposition of magma on the planet's surface, is driven by interior heat sources on a planet, including leftover heat from formation, radioactive decay, and tidal heating \citep[e.g.][]{Schubert.2001}. The magnitude of these internal heat sources depends on a planet's mass and age, among many other parameters. Given that the TRAPPIST-1 planets range in mass from 0.306 to 1.443 Earth masses \citep{Agol.2021}, and the age of the system is 7.6 $\pm$ 2.2 Gyr \citep{Burgasser.2017}, there is a wide range of possible internal heat values and corresponding magma emplacement rates. Further uncertainty is introduced due to our lack of constraints on the planets' interior compositions and structures.

Earth, Mars, and Io are known to have significantly different interior heat sources and planetary processes, and studies of these bodies have yielded a diverse range of magma emplacement rates \citep[see][and references within]{Lourenco.2018} that may provide useful benchmarks when considering the TRAPPIST-1 planets. Earth has plate tectonics, with most magma generated at plate boundaries. Mars is $\sim10$\% of Earth's mass, has stagnant lid tectonics, and is far less volcanically active. Io is $\sim1$\% of Earth's mass, yet it is more volcanically active because it is heated by tidal interactions with Jupiter. These 3 bodies have estimated magma emplacement rates of $Q_{MARS} = 0.06 \pm 0.3$ km$^{3}$ yr$^{-1}$ \citep{Hu.2022,Thomas.2023}, $Q_{EARTH} = 30 \pm 4$ km$^{3}$ yr$^{-1}$ \citep{Crisp.1984}, and an upper limit of $Q_{IO} = 413$ km$^{3}$ yr$^{-1}$ \citep{Breuer.2022} assuming magma density of 2900 kg m$^{-3}$, consistent with MORBs \citep[][Ch. 5]{Sigurdsson.2015} in all three cases. 

To account for as many plausible scenarios as possible, we sample a range of values for magma emplacement rate ($Q$) for the TRAPPIST-1 planets with an upper limit of 8500 km$^{3}$ yr$^{-1}$ set by \citet{Krissansen-Totton.2022}, which is an order of magnitude higher than Io's, and lower limits of 0.01 km$^{3}$ yr$^{-1}$, which is lower than modern Mars but on the same order of magnitude (Table \ref{t1_params}). \citet{Krissansen-Totton.2022} used a planetary evolution model to estimate the modern day magma emplacement rates on the TRAPPIST-1 planets over a wide range of model parameters and evolutionary scenarios. Their interior thermal evolution is calculated considering radioactive decay, heat from the core, convective and advective heat transport to the surface, and the impact of overlying rock on the melt boundary. Tidal heating is implicitly included by exploring a large range of initial radionuclide abundances, from 1/3x to 30x Earth's initial endowment. \citet{Barr.2018} reports that TRAPPIST-1 b and c may experience a sustained tidal heat flux of up to 4.01 and 1.62 W m$^{-1}$, respectively, while \citet{Dobos.2019} find  lower tidal heat fluxes of 1.97 and 1.04 W m$^{-1}$. These upper limits on tidal heat flux are comparable to or less than Io \citep[up to 4 W m$^{-1}$;][]{Breuer.2022}, and our parameter sweep easily encompasses them, as we explore $Q$ values over an order of magnitude larger than Io's for both TRAPPIST-1 b and c. Tidal heating is not expected to be significant on the other planets \citep[i.e., $<0.4$ W m$^{-1}$;][]{Dobos.2019}, yet the maximum explored $Q$ values all exceed Io's. On the other hand, it is possible that the TRAPPIST-1 planets are effectively geologically dead, like Mars, due to cooling over their long lifetimes.

\subsubsection{Equilibrium magmatic outgassing}

To complete our calculation of water outgassing rates, we calculate the expected outgassing from a magma parcel as it rises to the surface and equilibrates with the atmosphere. We make this calculation with the equilibrium magma outgassing chemistry model of \citet{Wogan.2020}. This model was developed for mantle-sourced volcanoes and will degas what is exsolved from the magma (i.e. what is above the saturation pressure). This model solves for gas-gas and gas-melt equilibrium in the \ch{C}-\ch{O}-\ch{H} system, including \ch{H2O}, \ch{CO2}, \ch{H2}, \ch{CO}, and \ch{CH4}. It has been validated against the outgassing models of \citet{Gaillard.2014}, \citet{Liggins.2020}, and \citet{Ortenzi.2020}.

We follow \citet{Wogan.2020} and assume values for input parameters based on the variety of conditions in the solar system, shown in Table \ref{og_params}. The inputs for this model are the magma temperature ($T$), the pressure at the surface ($P_{surf}$), the magma water content ($m_{\ch{H2O}}^{melt}$), the magma \ch{CO2} content ($m_{\ch{CO2}}^{melt}$), and the magma oxygen fugacity ($f_{\ch{O2}}$). The outputs are the gas production rates ($q_i$) for each species $i$ in moles of gas per kg of magma. 

Finally, the outgassing rate of any species $i$ to the atmosphere ($F_{i}$) is then, following \citet{Wogan.2020}:
\begin{equation} \label{eq:og_rate}
F_{i} = q_i \times Q \times \rho_{\rm mag}
\end{equation}
where we assume $\rho_{\rm mag} = 2900$ kg m$^{-3}$ is the density of the magma, consistent with Earth's oceanic crust \citep[][Ch. 5]{Sigurdsson.2015}, and $F_{i}$ is in units of moles yr$^{-1}$.

\subsection{Filters}\label{sec:filters}

After applying the theoretical model to the TRAPPIST-1 planets and exploring a large parameter space, we narrow the space of plausible model results by applying two filters based on current observational and plausible geochemical constraints.

\subsubsection{\ch{H2} escape filter}

With this filter, we use the constraint that telescope observations and atmospheric models find no evidence for large, \ch{H2}-dominated atmospheres on the TRAPPIST-1 planets  \citep{Wit.2018,Wakeford.2019,Moran.2018,Hori.2020,Turbet.2020} to identify model results with allowable \ch{H2} outgassing rates. To be consistent with the lack of observed \ch{H2}-dominated atmospheres, \ch{H2} must either not be the dominant outgassed species, or if it is, then  its  outgassing rate cannot exceed its atmospheric escape rate, otherwise \ch{H2} would build up in the atmosphere.   For all our model runs, we predict both the rate of \ch{H2O} and \ch{H2} outgassing, because we consider the equilibrium chemistry of the C-O-H system \citep{Wogan.2020}. We then apply our filter to  eliminate any model run where the \ch{H2} is the dominant outgassed species (by number of moles, not mass) and the \ch{H2} outgassing rate is higher than the estimated \ch{H2} escape rate (calculated below).

For our escape calculations we assume that hydrogen escape is energy-limited, which allows us to keep as many plausible scenarios as possible. Energy-limited escape is the theoretical upper limit on hydrogen escape as it assumes the limiting factor is the X-ray and ultraviolet (XUV) radiation from the star \citep[e.g., see][Ch. 5]{Catling.2017}. We calculate the energy-limited escape rate of hydrogen ($\Phi_{\ch{H}}$) on the TRAPPIST-1 planets according to \citet{Chassefiere.1996} and \citet{Luger.2015}:

\begin{equation}
\Phi_{\ch{H}} = \frac{\epsilon_{\rm XUV} F_{\rm XUV} r_p}{4GM_pK_{\rm tide}m_{\ch{H}}}
\label{h2_escape}
\end{equation}

where $\epsilon_{\rm XUV}$ is the escape efficiency factor equal to the portion of XUV radiation that contributes to escape, assumed to be 0.3, which is the upper end of what is expected in hydrogen-rich atmospheres \citep{Chassefiere.1996, Wordsworth.2013}. $F_{\rm XUV}$ is the received XUV flux for each planet from \citet{Becker.2020}, shown in Table \ref{t1_params}. $r_p$ and $M_p$ are the radius and mass of the planet, respectively, from \citet{Agol.2021}. $G$ is the gravitational constant. $K_{\rm tide}$ is a tidal correction term assumed to be 1 for simplicity. The mass of hydrogen is $m_H$. The hydrogen escape rates from this equation for the TRAPPIST-1 planets are shown in Table \ref{t1_params}.

\subsubsection{Evolutionary filter}

With this filter, we identify the most plausible scenarios by requiring that the TRAPPIST-1 planets do not exceed the 1 wt\% upper limit on $m_{\ch{H2O}}^{mantle}$ (see Section \ref{sec:interior_water}) throughout their history. We do this by extrapolating our estimated modern water outgassing rate values ($F_{\ch{H2O}}$) back in time, adding the outgassed water back into the mantle, and then calculating the value of $m_{\ch{H2O}}^{mantle}$ over the history of the planets. 

When extrapolating the outgassing rate and mantle water storage capacity back in time, we are forced to make several assumptions about water partitioning and recycling to simplify our approach. First, we assume that there is no recycling of water from the exterior back into the mantle (e.g., via plate tectonics), consistent with the rest of our model calculations and the lack of evidence for subduction on rocky bodies other than Earth \citep{Stern.2018axqs}. Second, we assume that the TRAPPIST-1 planets have not had 10-100 km deep surface oceans (Earth's ocean is $\sim$3.6 km average depth) that result in the 0.1-1 GPa overburden pressure required to suppress volcanic outgassing for the majority of their history \citep[given Earth-like gravity and ocean crust composition;][]{Krissansen-Totton.2021}. Third, we assume that the maximum mantle water content is always 1 wt\%, although internal water storage changes over time as the planets cool \citep[e.g.,][]{Dong.2021}. However, the 1 wt\% upper limit is still valid since it is the maximum possible water storage, not the specific water storage over time for each planet. Fourth, we assume that $F_{\ch{H2O}}$ is constant throughout the planets' histories, and equal to our calculation of the modern value. In reality, it is likely that $F_{\ch{H2O}}$ would have been higher in the past due to the early planet having a higher internal heat flux from formation and radionuclide decay.

We made the above assumptions so that the evolutionary filter is as permissive as possible, opting to keep model runs instead of eliminating them given our lack of constraints on the history of the TRAPPIST-1 planets. Thus, changing the above assumptions would generally make this filter stricter. For example, if we instead assumed that $F_{\ch{H2O}}$ was higher in the past, it would mean more water must be added back into the mantle in the past, the 1 wt\% upper limit is more likely to be violated, and more model runs would be eliminated. Past water recycling would likely cause higher $F_{\ch{H2O}}$ in the past as well, leading to the same effect as above. However, this effect may be partially offset because the ingassed water would simultaneously be removed from the mantle as we extrapolate back in time. This filter would also be stricter if we modeled changing mantle water capacity over time instead of setting a maximum mantle water content. This change would result in a lower mantle water content limit and more model runs eliminated for violating it. Our assumption of the ocean overburden pressure has the opposite effect, as it would result in lower outgassing rates in the past, which would eliminate fewer model runs than with our current assumption. These assumptions are therefore offsetting, but show that we are being as permissive as possible while maintaining model simplicity.

To calculate the mantle water mass fraction over time for each planet, $m_{\ch{H2O}}^{mantle}(t)$, we use conservation of mass to derive the following equation:
\begin{equation} \label{wmf_evol}
m_{\ch{H2O}}^{mantle}(t) = \frac{M_{\rm mantle}(0) m_{\ch{H2O}}^{mantle}(0) + F_{\ch{H2O}}t}{M_{\rm mantle}(0) + F_{\ch{H2O}}t}
\end{equation}
where $t$ is the time before present, with $t=0$ denoting the present day, so $M_{\rm mantle}(0)$ is the modern mass of the mantle and $M_{\rm mantle}(t) = M_{\rm mantle}(0) + F_{\ch{H2O}}t$. For each model run, for each planet, we can invert Equation \ref{wmf_evol} to derive a maximum allowable water outgassing rate, $F^{\max}_{\ch{H2O}}$, such that the 1 wt\% upper limit on $m_{\ch{H2O}}^{mantle}$ is not exceeded throughout the planet's history:
\begin{equation} \label{filter2_eq}
    F^{\max}_{\ch{H2O}} = \frac{M_{\rm mantle}(0)[0.01 - m_{\ch{H2O}}^{mantle} (0)]}{0.99 \times t}
\end{equation}
where we have substituted 0.01 = 1 wt\% for $m_{\ch{H2O}}^{mantle}(t)$. 

To eliminate the fewest possible scenarios, we choose values for the age of the system and the core mass fraction that maximize $F^{\max}_{\ch{H2O}}$. First, $M_{\rm mantle}$ is calculated assuming a small core mass fraction of 18 wt\%, consistent with other interior modeling studies of TRAPPIST-1 \citep[e.g.,][]{Agol.2021}. Second, we use the minimum age of the TRAPPIST-1 system, 5.4 Gyrs \citep{Burgasser.2017}, as the largest value for $t$. For TRAPPIST-1 c, these parameter values result in  $F^{\max}_{\ch{H2O}}$ values up to $1.2 \times 10^{13}$ kg yr$^{-1}$ depending on $m_{\ch{H2O}}^{mantle}$ in a given model run. In contrast, selecting the median age of the system (7.6 Gyr) and an Earth-like core mass fraction (32.5 wt\%) would result in $F^{\max}_{\ch{H2O}}$ values up to $7 \times 10^{12}$ kg yr$^{-1}$, which would make this filter stricter.

\section{Results}

Here we describe the results of experiments that explore a broad range of parameters for the seven TRAPPIST-1 planets to generate plausible ranges for planetary water content, magma emplacement rates, and outgassing rates for water and other key gases. We then provide results for specific TRAPPIST-1 c test cases with fixed magma emplacement rates corresponding to those on Mars, Earth, and Io. 

\subsection{Parameter correlations and sensitivity to filters}

We first ran our model 100,000 times with Monte Carlo sampling to explore the plausible ranges for mantle water content, magma emplacement, and outgassing for the TRAPPIST-1 planets. Here, we sampled the full parameter ranges in Table \ref{og_params}, with the volumetric magma emplacement rate ($Q$) uniformly sampled in base 10 log space within upper limits given in Table \ref{t1_params}. We then applied both of the filters described in Section \ref{sec:filters} to remove cases that violated observational and geochemical constraints. This process is similar across the seven planets because most of the explored parameter ranges are the same, except for $Q$, which is different for each planet because it depends on mass and radius. Planetary mass and radius also factor into calculation of the energy-limited \ch{H2} escape rate filter, and the mantle mass and mantle water content in the evolutionary filter. We also tested two additional scenarios where the maximum mantle water content is 0.5 wt\% and 3 wt\%, as opposed to the nominal 1 wt\%. These results thus represent all of the allowed scenarios on the TRAPPIST-1 planets according to our modeling framework.

\begin{figure*}[htbp]
    \centering
    \includegraphics[width=\textwidth]{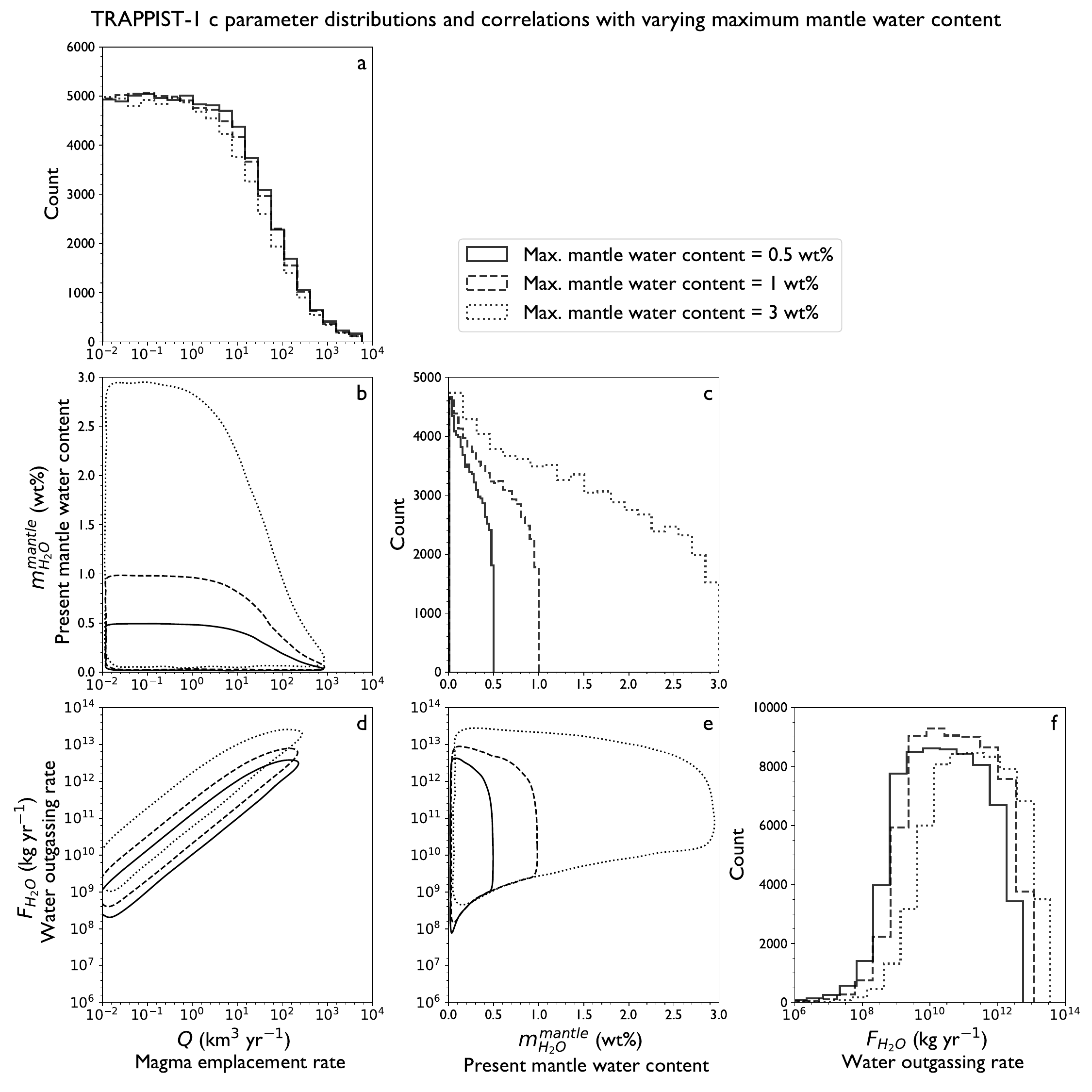}
    \caption{Parameter distributions and correlations with varying maximum mantle water content for TRAPPIST-1 c. Plots are derived from 100,000 model runs where the full parameter sweep is executed then the two filters are applied to eliminate model runs, described in Section \ref{sec:filters}. (a, c, f) Posterior parameter distributions. (b, d, e) Parameter correlations. The contours contain 95\% of the model runs, which are homogeneously distributed within.}
    \label{t1c_correlations}
\end{figure*}

Figure \ref{t1c_correlations} shows the posterior distributions and correlations of the mantle water content ($m_{\ch{H2O}}^{mantle}$), magma emplacement rate ($Q$), and water outgassing rate ($F_{\ch{H2O}}$) for TRAPPIST-1 c, revealing intrinsic relationships between these parameters and the impact of the two filters on the allowed phase space. First, we consider the nominal model scenario where the maximum mantle water content is 1 wt\%. Figure \ref{t1c_correlations}b shows the envelope of constraints imposed by the evolutionary filter, where only lower outgassing rates, and thus lower $Q$, are possible for a planet with a very wet mantle at present day (as higher rates would imply mantle water contents higher than 1 wt\% in the past). Higher $Q$ values are thus only permitted for drier present interiors. Figure \ref{t1c_correlations}d shows that $Q$ strongly correlates with $F_{\ch{H2O}}$, which is expected because it is a multiplicative factor and thus $F_{\ch{H2O}}$ is directly proportional to $Q$ (Eq. \ref{eq:og_rate}). Additionally, the \ch{H2} escape filter eliminates model runs with high $Q$ because higher $Q$ values increase $F_{\ch{H2}}$, which cannot be higher than the energy-limited \ch{H2} escape rate in order to be consistent with observational constraints. Figure \ref{t1c_correlations}e also shows constraints from the evolutionary filter, with a lower upper boundary on $F_{\ch{H2O}}$ when present $m_{\ch{H2O}}^{mantle}$ is higher - again because $Q$ must be lower when mantle water content is high. Although Figure \ref{t1c_correlations} shows correlations for TRAPPIST-1 c only, we find that these behaviors broadly hold for all of the TRAPPIST-1 planets.

Figure \ref{t1c_correlations} also shows that our model results are only slightly sensitive to varying the theoretical maximum mantle water content, partially because the evolutionary filter eliminates model runs with high present mantle water content, motivating us to use the nominal 1 wt\% upper limit going forward. The maximum and median water outgassing rates are 3x and 3.8x higher, respectively, when the maximum mantle water content is 3 wt\% compared to 1 wt\%; this difference is small relative to the $\sim6$ orders of magnitude that the water outgassing rate distributions span (Figure \ref{t1c_correlations}f). The evolutionary filter, which was modified to reflect the varying maximum mantle water content, partially explains this behavior because it is more likely to eliminate scenarios with present mantle water content close to the theoretical maximum. Therefore, the impact of lowering the theoretical maximum is buffered because model runs near the theoretical maximum are already eliminated.

In the 1 wt\% upper limit model runs, the two filters combine to eliminate 29-37\% of model runs across the seven TRAPPIST-1 planets, where the vast majority are eliminated by the evolutionary filter. On one hand, the \ch{H2}-escape filter only eliminates $\sim1$\% of model runs, showing that \ch{H2} outgassing rates are unlikely to exceed the energy-limited \ch{H2} escape rates in the broad parameter space we searched. On the other hand, the evolutionary filter eliminates up to to 37\% of model runs, showing that many plausible parameter combinations for the modern day TRAPPIST-1 planets are inconsistent with even the most permissive of evolutionary considerations.

\subsection{Posterior distributions for mantle water content, magma emplacement, and outgassing on the TRAPPIST-1 planets}

\begin{figure*}[htbp]
    \centering
    \includegraphics[width=0.7\textwidth]{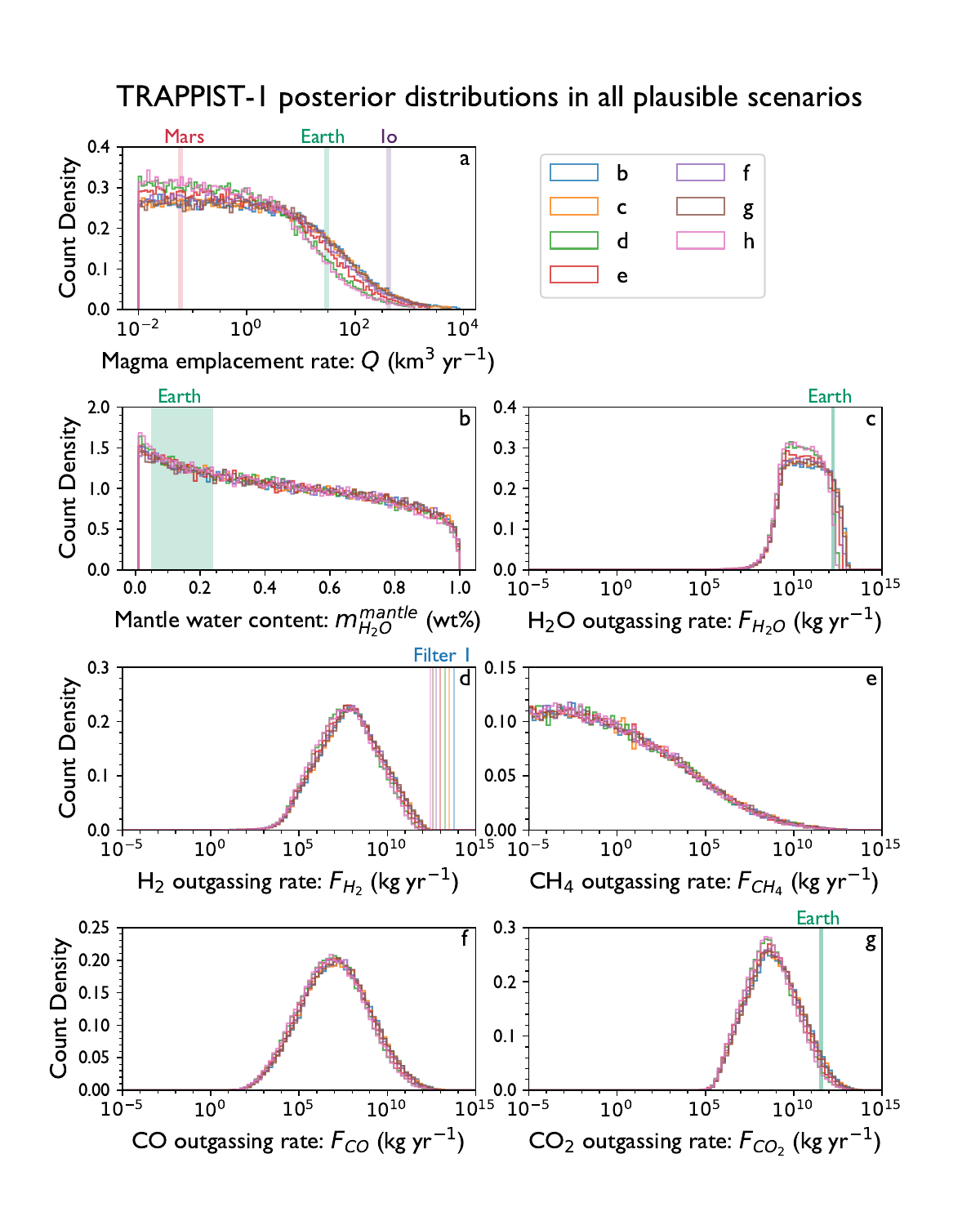}
    \caption{Posterior distributions of model values for all TRAPPIST-1 planets in all plausible model scenarios. Plausible model scenarios are derived by first executing the full parameter sweep and then applying empirical and theoretical filters to eliminate model runs, described in Section \ref{sec:filters}. The most likely scenario on the TRAPPIST-1 planets is that they have (a) low magma emplacement rates, (b) present mantle water content consistent with Earth or lower (although wetter mantles are possible), and (c) a water outgassing rate lower than Earth's (although it can be up to an order of magnitude higher). Vertical bars are values from the solar system or filter application: (a) $Q$ for Mars, Earth, and Io are 0.06, 30, and 413 km$^3$ yr$^{-1}$, respectively (Section \ref{Q_methods}). (b) Earth $m_{\ch{H2O}}^{mantle}$ is $0.049-0.24$ wt\% \citep{Ohtani.2020}. (c, g) Earth's \ch{H2O} and \ch{CO2} outgassing rates are $1.35 \times 10^{12}$ - $2.07 \times 10^{12}$ kg yr$^{-1}$ and $2.86 \times 10^{11}$ - $4.62 \times 10^{11}$ kg yr$^{-1}$, respectively \citep[][p. 203]{Catling.2017}. (d) Energy-limited \ch{H2} escape rates from filter 1 (Section \ref{sec:filters}).}
    \label{t1_posteriors}
\end{figure*}

Figure \ref{t1_posteriors} shows the posterior distributions for all seven TRAPPIST-1 planets for model parameters and outgassing rates after the Monte Carlo sampling and application of the two filters. Figure \ref{t1_posteriors}a shows that low magma emplacement rates are more likely for all TRAPPIST-1 planets. The median magma emplacement rate for TRAPPIST-1 c is 0.8 km$^{3}$ yr$^{-1}$, which is 2 orders of magnitude lower than Earth's and indicates Mars-like magma emplacement is more likely. Additionally, the distributions are approximately uniform below 1 km$^{3}$ yr$^{-1}$, suggesting that all values below this threshold are equally likely. The count density drops off steeply for high values of $Q$, where there is only a 13.5\% chance that a model run will have $Q$ higher than Earth and a 1.96\% chance it will have $Q$ higher than Io. Finally, the distributions for TRAPPIST-1 d (green line) and h (pink line) are shifted toward lower values relative to the other planets because they have lower masses, and thus lower upper limits on $Q$, compared to the rest of the planets. 
 
Figure \ref{t1_posteriors}b shows a preference towards the low end of the explored range for mantle water content, although values throughout the entire range are plausible for all TRAPPIST-1 planets. For example, the median mantle water content for TRAPPIST-1 c is 0.42 wt\%, and the most frequent values are those at or below Earth's range of estimates ($0.049-0.24$ wt\% \citep{Ohtani.2020}). Although there is a preference for lower values, there are still many model scenarios with wetter mantles, as the count density does not go to zero until the very top of the explored range is reached. There are currently no measurements of the mantle water content on Mars or Io to compare to.

Despite having mantle water contents comparable to or even higher than Earth's, Figure \ref{t1_posteriors}c shows that the water outgassing rates for the TRAPPIST-1 planets are likely less than Earth's. The outgassing rates follow an approximate log-normal distribution centered an order of magnitude below Earth's value, but with upper limits that exceed it. For example, the median value for TRAPPIST-1 c is $5 \times 10^{10}$ kg yr$^{-1}$, which is over an order of magnitude below Earth's value of $\sim2 \times 10^{12}$ kg yr$^{-1}$. This disparity is likely because (1) Earth's water outgassing rate is enhanced by recycling of surface water during plate-tectonics-induced subduction processes and (2) we sample uniform priors for e.g. magma emplacement rate with much of the distribution below Earth's value. In comparison, our calculations only assume mantle-derived magmas that are not impacted by subduction and transport of surface water. The modeled upper limits on water outgassing are $2.0 - 12.4 \times 10^{12}$ kg yr$^{-1}$, which at the higher end are an order of magnitude higher than Earth's value. Once again, the distributions for TRAPPIST-1 d and h are shifted toward lower or intermediate values of water outgassing rate because they have lower masses, lower upper limits on $Q$, and thus lower upper limits on $F_{\ch{H2O}}$ compared to the rest of the planets. 

Figures \ref{t1_posteriors}d-g show the outgassing rates of other gases that are self-consistently calculated in our model along with the water outgassing rate. The \ch{H2}, \ch{CO}, and \ch{CO2} outgassing rates follow approximate log-normal distributions that are shifted toward lower values relative to the water outgassing rate. The \ch{CO2} outgassing rate is most frequently lower than Earth's because, although the explored \ch{CO2} magma concentration is consistent with what is found on Earth, $Q$ is preferred to be lower than Earth's value in the corresponding posterior distribution in Figure \ref{t1_posteriors}a. The \ch{CH4} outgassing rate is much lower than all of the others because it requires a very reduced magma (i.e., low oxygen fugacity) to be the dominant species. Posterior distributions of other model parameters are omitted here because they did not change significantly from their flat prior distributions, but are shown in Appendix A. Posterior distributions in flux units are shown in Appendix B.

\subsection{Solar system cases: Earth, Mars, and Io}

\begin{figure*}[htbp]
    \centering
    \includegraphics[width=0.8\textwidth]{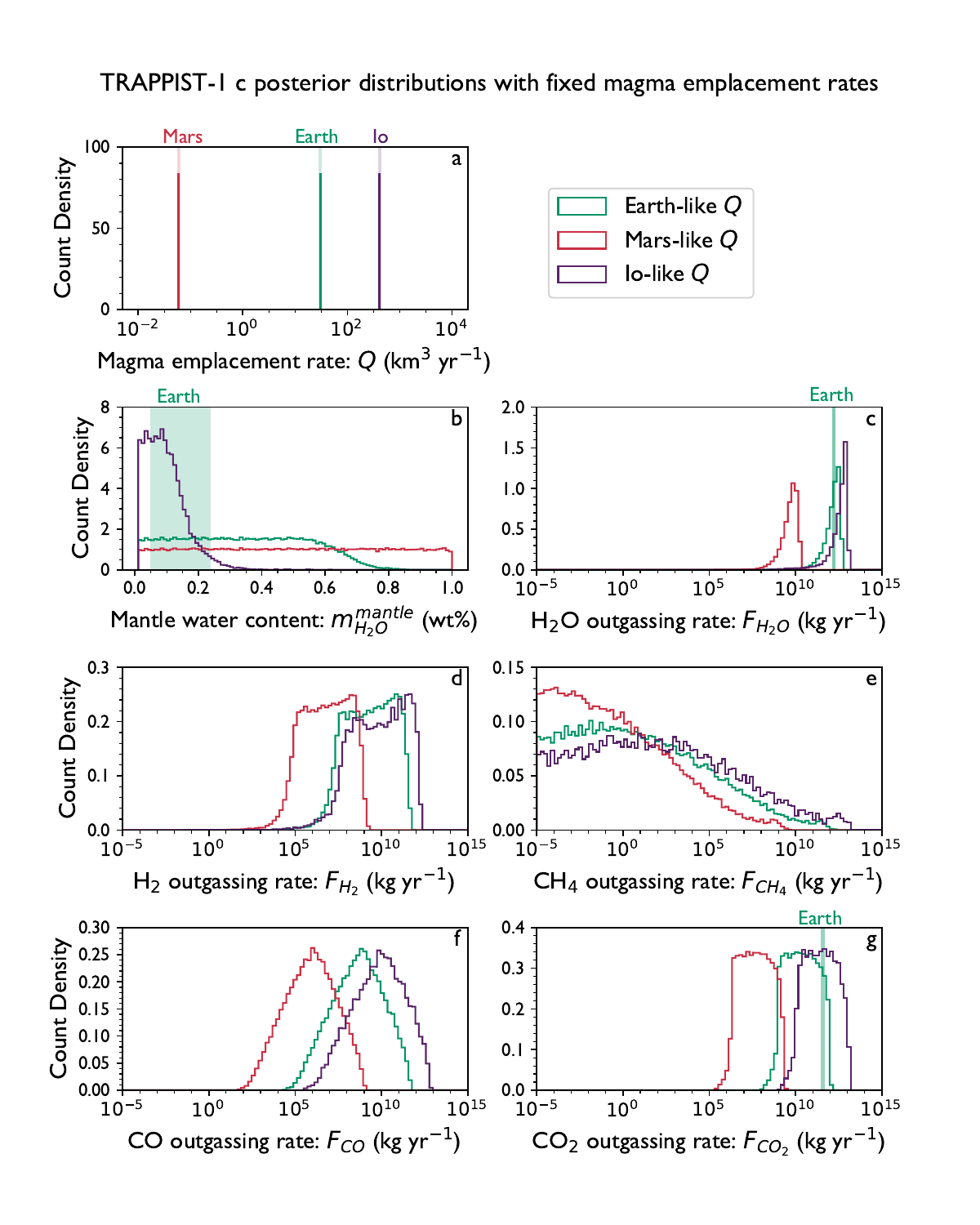}
    \caption{Posterior distributions of model values for TRAPPIST-1 c in plausible model scenarios with fixed magma emplacement rates based on those of Earth, Mars, and Io. Plausible model scenarios are derived by first executing the full parameter sweep but with $Q$ fixed to one of the solar system values, then applying empirical and theoretical filters to eliminate model runs that violate known empirical and theoretical constraints, as described in Section \ref{sec:filters}. In this case we apply the filters using the mass and radius of TRAPPIST-1 c as a representative case. Vertical bars are values from the solar system: (a) $Q$ for Mars, Earth, and Io are 0.06, 30, and 413 km$^3$ yr$^{-1}$, respectively (Section \ref{Q_methods}). (b) Earth $m_{\ch{H2O}}^{mantle}$ is $0.049-0.24$ wt\% \citep{Ohtani.2020}. (c, g) Earth's \ch{H2O} and \ch{CO2} outgassing rates are $1.35 \times 10^{12}$ - $2.07 \times 10^{12}$ kg yr$^{-1}$ and $2.86 \times 10^{11}$ - $4.62 \times 10^{11}$ kg yr$^{-1}$, respectively \citep[][p. 203]{Catling.2017}.}
    \label{t1c_fixedQ_posteriors}
\end{figure*}

While the above simulations showed the broad parameter sweep of allowable cases, we also explored the range of possible outcomes for specific cases with fixed magma emplacement rates ($Q$) spanning known values for solar system bodies (see Section \ref{Q_methods}). Figure \ref{t1c_fixedQ_posteriors} shows the results of running our model 100,000 times while Monte Carlo sampling the parameter ranges in Table \ref{og_params}, with $Q$ fixed at representative values for Earth, Mars, and Io (Section \ref{Q_methods}), and then applying both of the filters described in Section \ref{sec:filters}. We show the results of applying this treatment to TRAPPIST-1 c as a representative scenario, noting that in reality the magma emplacement rate should scale with planet mass. Because the magma emplacement rates are fixed, they no longer incorporate each planet's mass and radius, which will now only factor into calculation of the filters.  

The three fixed $Q$ values, and the filters, strongly constrain the inferred  $m_{\ch{H2O}}^{mantle}$ and $F_{\ch{H2O}}$ in these model runs (Figures \ref{t1c_fixedQ_posteriors}b and c). With a highly volcanic, Io-like $Q$, TRAPPIST-1 c would likely have $m_{\ch{H2O}}^{mantle}$ at or less than that of Earth but $F_{\ch{H2O}}$ over an order of magnitude larger than Earth. On the other hand, if TRAPPIST-1 c is volcanically dead and has a low, Mars-like $Q$, then there are no constraints on $m_{\ch{H2O}}^{mantle}$ within its sampled range, as all values are equally likely, and $F_{\ch{H2O}}$ would be consistently orders of magnitude lower than Earth's water outgassing rate. The Earth-like $Q$ case provides a middle scenario which favors high $F_{\ch{H2O}}$ but many possible values of $m_{\ch{H2O}}^{mantle}$ below 0.6 wt\%. The median values of $F_{\ch{H2O}}$ for TRAPPIST-1 c with Io-like, Earth-like, and Mars-like $Q$ are 4.6, 1.7, and $0.005 \times 10^{12}$ kg yr$^{-1}$, respectively.

The outgassing rates for the additional molecules considered here are proportional to $Q$ (Figures \ref{t1c_fixedQ_posteriors}d-g), where an Io-like $Q$ causes higher outgassing rates for all species relative to a Mars-like $Q$. The \ch{H2}, \ch{CO}, and \ch{CO2} outgassing rates follow approximately log-normal distributions that are offset from each other based on the difference in $Q$. The \ch{CH4} outgassing rate skews toward low values in all cases, but the offset based on $Q$ is still present. Posterior distributions of other model parameters in these solar system test cases are omitted here because they did not change significantly from their flat prior distributions, but are shown in Appendix A. Posterior distributions in flux units are shown in Appendix B.

\section{Discussion}

We explored a comprehensive and broad parameter space to place plausible boundaries on outgassing rates for water and other key gases for the TRAPPIST-1 planets to support interpretation of data from upcoming observations of these worlds, and others. Although our explored parameter space covers a wide range of chemical and physical conditions, we constrained this space by using observational evidence for a lack of \ch{H2}-dominated atmospheres (\ch{H2}-escape filter) and theoretical limits on the maximum mantle water mass fraction over time (evolutionary filter).

Our results indicate that drier mantles are preferred within the broader explored range of mantle water content ($m_{\ch{H2O}}^{mantle} = $ 0-1 wt\%; Figure \ref{t1_posteriors}b). This arises due to our assumption that the TRAPPIST-1 planets have terrestrial interiors with mantle water contents that remain below a 1 wt\% upper limit throughout the $\geq$5.4 Gyr age of the TRAPPIST-1 system \citep{Burgasser.2017}. The preference for lower $m_{\ch{H2O}}^{mantle}$ values is more consistent with Earth's mantle water content \citep[$0.049-0.24$ wt\%;][]{Ohtani.2020}. Although higher $m_{\ch{H2O}}^{mantle}$ values are still possible, there is no known terrestrial analog for comparison.

Our model results for magma emplacement rates also indicate that the TRAPPIST-1 planets are currently more likely to have low-to-no volcanic activity. Higher volcanic activity is disfavored as it often implies past mantle water content over the 1 wt\% upper limit. For all seven planets, the posterior distributions for magma emplacement rates ($Q$) skew toward the low end of their allowed ranges, disfavoring rates that may be associated with high tidal heat fluxes (Figure \ref{t1_posteriors}a). For example, 86.5\% of plausible model runs for TRAPPIST-1 c have $Q$ below Earth's value \citep[30 km$^{3}$ yr$^{-1}$;][]{Crisp.1984}, and 52.5\% are below 1 km$^{3}$ yr$^{-1}$, indicating that extremely low, Mars-like rates of volcanism are the most likely. Conceptually, extremely low volcanic activity on the TRAPPIST-1 planets would be consistent with the fact that they are Earth-sized, and at least $\sim$1 Gyr older than the comparable solar system bodies, so there has been more time for their planetary interiors to cool. 

Our results indicate that the water outgassing rates on the TRAPPIST-1 planets are more likely to be lower than Earth's, but the plausible range also includes outgassing rates that are an order of magnitude higher than Earth's. Lower outgassing rates are more likely for two reasons: (1) we consider mantle-derived magmas in this study, which have lower water content than magmas derived from the recycling of surface material, like arc magmas on Earth, which acquire high water content due to subduction of hydrated oceanic crust, and (2) the highest modeled water outgassing rates are removed by the requirement that the TRAPPIST-1 planets' mantle water content cannot exceed a 1 wt\% upper limit during the history of the system. This results in, for example, a median water outgassing rate for TRAPPIST-1 c of $5 \times 10^{10}$ kg yr$^{-1}$ ($\sim$0.03x Earth's). That being said, high water outgassing rates up to $1.2 \times 10^{13}$ kg yr$^{-1}$ ($\sim$8x Earth's) can be achieved for a broad range of mantle water contents if the TRAPPIST-1 planets have high magma emplacement rates (Figure \ref{t1c_correlations}).

Water is most frequently the dominant outgassed species in our model results and indicates that if secondary atmospheres are present, they will likely be dominated by water and/or its photochemical byproducts. However, there are still model scenarios where \ch{CO2}, \ch{CO}, and \ch{H2} outgassing rates exceed water outgassing rates. The outgassing rate distributions and self-consistent scenarios can be used in future modeling studies to explore possible secondary atmospheres consistent with outgassing. Our calculated outgassing rates are provided in an online repository and available as fluxes in mol s$^{-1}$ m$^{-2}$ units in Appendix B (Figures \ref{appendix_posteriors_fluxunits} and \ref{appendix_solarcases_fluxunits}).

\subsection{Comparison to other models of water outgassing and planetary evolution}

We have grounded our modeling approach in present-day observational constraints and knowledge of the solar system to statistically search a broad parameter space of possible present-day scenarios on the TRAPPIST-1 planets. Given the lack of constraints on the geological properties of these planets, a huge variety of present-day interior scenarios is possible. With our modeling approach, we can easily test a wide range of these present-day scenarios and provide information that informs observations. This approach is complemented by planetary evolution models, which consider the history of the system when predicting the present state. Many previous evolutionary models have investigated similar topics for the TRAPPIST-1 planets specifically \citep[e.g.,][]{Bourrier.2017, Barth.2021, Krissansen-Totton.2022} and terrestrial planets orbiting M stars broadly \citep[e.g.,][]{Moore.2020, Moore.2023, Moore.2024}, which represent a sample of the many possible trajectories of the system. We do consider the evolution of the TRAPPIST-1 planets with our evolutionary filter (discussed below); however, the main strength of our modeling approach is the flexible statistical method we have used to search a broad parameter space of present-day properties.

We have constructed our evolutionary filter to be simple and permissive so that it is consistent with the lack of constraints on the history of the system. Given the lack of data, evolutionary models are forced into assumptions about the planets' pasts that, while reasonable, may not be true. For example, the only data constraining the interiors of the TRAPPIST-1 planets are the observed mass and radius \citep{Grimm.2018, Agol.2021}, and estimates of elemental abundances based on the local stellar composition \citep[e.g., Fe/Mg;][]{Unterborn.2018}. Others have estimated interior properties from these data \citep[e.g.,][]{Agol.2021}, but there are many interior scenarios consistent with the data at present-day, and it is highly uncertain how many of these interior properties would have changed in the past. Therefore, we have taken the simplest approach when designing the evolutionary filter so that a wide variety of reasonable scenarios are not ruled out. As described in Methods, our evolutionary assumptions are as permissive as possible while maintaining simplicity: (1) water outgassing is constant, (2) there is no water recycling or regassing, (3) there are no large oceans that can suppress outgassing, and (4) the maximum mantle water content is constant.

Our resultant distributions of water and methane outgassing rates encompass previous estimates that make a variety of different assumptions, highlighting how our approach has established reasonable statistical boundaries on these values. The whole-planet evolutionary model of \citet{Krissansen-Totton.2022} estimates that the modern water outgassing rate across all TRAPPIST-1 planets does not exceed $\sim10^{12}$ kg yr$^{-1}$, and \citet{Moore.2020} estimate that peak water outgassing for Earth-like planets around M stars is $\sim7.7 \times 10^{12}$ kg yr$^{-1}$, but occurs within the first billion years of the system. Both of these water outgassing rates are within our calculated boundaries, but were obtained using more detailed evolutionary models with different assumptions; for example, both models assume some form of plate tectonics occurred and that interior mineralogy is broadly similar to Earth. Additionally, our calculated methane outgassing rates are consistent with a previous application of the magma degassing model \citep{Wogan.2020}, where the distribution of methane outgassing rates for Earth-like $Q$ in Figure \ref{t1c_fixedQ_posteriors}e is directly comparable to \citet{Wogan.2020} Figure 4d.

The 1 wt\% upper limit on mantle water content we explore is geochemically plausible, but is much higher than, the mantle water content derived from current magma ocean models. The 1 wt\% water content in the TRAPPIST-1 planets' mantles converts to up to $\sim50$ Earth oceans depending on planet size and core mass fraction. Our model does not address how water is incorporated into the mantle during formation, but previous models that calculate the partitioning of water in planet interiors have not obtained the high mantle water content we test here. For example, \citet{Barth.2021} find that up to 5 Earth oceans can be sequestered in the mantles of the outer TRAPPIST-1 planets. \citet{Moore.2023} find that up to 6 Earth oceans can be sequestered in Earth-mass planet mantles, but possibly up to the saturation limit of 12 Earth oceans for extreme parameters given their assumed mineralogy. Other studies assume up to 12-15 Earth oceans of storage in terrestrial planet mantles is possible based on present-day Earth's mantle water capacity \citep{Cowan.2014, Komacek.2016}. Future theoretical and observational work that explores mechanisms for obtaining the theoretical maximum water content will thus bear on our model results going forward. However, the highest water outgassing rates in our models occur for lower water mantle water content, so the effect on water outgassing may be minimal.

\subsection{Constraints on internal structure and composition of the TRAPPIST-1 planets}

Our model's predicted relationship between lower mantle water content and higher outgassing rates provides a mechanistic explanation for the possible distribution of mass between the inner TRAPPIST-1 planets' (b, c, and d) atmospheres and interiors. If there is exterior water on the inner planets, it will most likely be in the vapor phase as an atmosphere \citep{Turbet.2019,Turbet.20207e,Agol.2021}. To match the observed total planetary density, the low-density of the atmosphere would have to be offset by a denser solid planet, when compared to an atmosphere-free case. One way to achieve the denser solid body is for the planetary interior, including the mantle, to have a lower water content with more high-density nominally anhydrous minerals. This argument applies in reverse too, where if there is a tenuous water-vapor-containing atmosphere, or none at all, the solid planet would have to be relatively less dense and could thus have a higher interior water content with more low-density hydrous minerals. These scenarios are consistent with our results as the denser modeled scenarios with low mantle water content have the highest magma emplacement rate and highest water outgassing rates, which could theoretically support the thicker water-vapor-containing atmosphere against atmospheric escape. Applying this argument in reverse, we also find that a lower density mantle with the highest mantle water content would have lower magma emplacement rates and lower water outgassing rates, which would be consistent with a tenuous or no water-vapor-containing atmosphere.

Our results also strongly disfavor present-day Io-like volcanism on the TRAPPIST-1 planets, suggesting that the tidal or induction heating required to maintain these high levels of volcanism may not currently be present. This result is inconsistent with some existing model predictions \citep{Kislyakova.2017,Barr.2018}. For example, \citet{Barr.2018} use an interior thermal model to predict that TRAPPIST-1 b has high volcanic activity with an interior heat flux twice as high as Io. According to our results, magma emplacement rates greater than or equal to Io's occur in only 1.8\% of model scenarios. In these model scenarios, we predict that the high magma emplacement rates lead to the highest possible water outgassing rate (see Figure \ref{t1c_fixedQ_posteriors} with Io-like $Q$). Therefore future constraints on the water outgassing rate for the TRAPPIST-1 planets via measurement of atmospheric water abundance have the potential to also constrain $Q$ (via the correlation in Figure \ref{t1c_correlations}b), and thus may have implications for the degree of tidal or induction heating on the TRAPPIST-1 planets.

\subsection{Outgassing as a source of atmospheres for the TRAPPIST-1 planets}

The outgassing rates of water and the other gases provided here can be coupled with estimates of atmospheric escape to assess the presence and observability of water in the TRAPPIST-1 planet atmospheres. 

We find that the predicted water outgassing rates in our plausible model scenarios can likely balance atmospheric escape, and could support stable water-vapor-containing atmospheres on the TRAPPIST-1 planets. For TRAPPIST-1 b, the closest planet to the parent star with the highest stellar-radiation-driven escape rates, we compare our modeled water outgassing rates to the modeled water escape rates of \citet{Krissansen-Totton.2022} and \citet{Gialluca.2024}. At 1 billion years after planet formation, both studies predict a median water escape rate of $\sim1.8 \times 10^{10}$ kg yr$^{-1}$ which decreases with time. This value is well below our upper limit for water outgassing of $1.2 \times 10^{13}$ kg yr$^{-1}$. Therefore, our modeled outgassing rates are potentially high enough to balance the water escape rates on TRAPPIST-1 b and the other planets, indicating that steady-state water containing atmospheres are possible. Future studies are needed to further evaluate and characterize these steady states by coupling the outgassing rates provided here to atmospheric chemistry and escape models.

In fact, the higher plausible \ch{H2O} outgassing rates could produce significant atmospheres, or form oceans, on the temperate TRAPPIST-1 planets. The upper limit on the \ch{H2O} outgassing rate for TRAPPIST-1 c is $1.2 \times 10^{13}$ kg yr$^{-1}$, producing $\sim50$ Earth oceans of degassed water over the minimum 5.4 Gyr age of the system. This amount of water can likely survive atmospheric escape \citep{Gialluca.2024} and could theoretically be responsible for surface water accumulation on the habitable zone planets.

Moreover, the predicted higher \ch{CO2} outgassing rates could also contribute to high pressure \ch{CO2} atmospheres. 2\% of our model runs for TRAPPIST-1 c have a \ch{CO2} outgassing rate over $10^{12}$ kg yr$^{-1}$. When extrapolated to the minimum age of the system, this would result in an atmosphere with 100s of bars of \ch{CO2} if no escape or recycling processes are considered. Deriving an upper limit on \ch{CO2} escape would provide further constraints on the models that would preclude excessive, and unobserved \citep{Greene.2023,Zieba.2023} \ch{CO2} build up. 

\subsection{Implications for current and future observations of TRAPPIST-1}

Our results do not preclude water vapor in the atmosphere of TRAPPIST-1 c, and suggest its plausibility. As described in the Introduction, the secondary eclipse observation of TRAPPIST-1 c is consistent with several water-vapor-containing atmospheres \citep{Zieba.2023, Lincowski.2023}. Our modeled outgassing rates can plausibly balance the high atmospheric escape rates on TRAPPIST-1 c, so this cannot be used as a basis to rule out the presence of a water-vapor-containing atmosphere and eliminate the scenarios presented in \citet{Lincowski.2023}. Instead, additional observations will be needed to further discriminate between possible atmospheric scenarios. 

For the TRAPPIST-1 habitable zone planets, a large sustained water outgassing rate throughout their evolution may permit habitability, even after early water loss. Significant water loss is expected in the TRAPPIST-1 system from high radiation during and after the pre-main sequence phase \citep[e.g.][]{Birky.2021, Gialluca.2024}, which can be augmented by intense stellar flares throughout the star's lifetime \citep[e.g.][]{Yamashiki.2019, Amaral.2022}. However, even early loss of any surface volatile inventory could be offset by outgassed water from the interior that accumulates on the surface over time. In this scenario, early catastrophic loss of an M dwarf planet's ocean and/or atmosphere does not necessarily preclude later habitability, and this could increase the chances that the TRAPPIST-1 habitable zone planets are indeed habitable \citep[e.g.][]{Moore.2020,Krissansen-Totton.2024}.

Our results also demonstrate that observations of TRAPPIST-1 planetary atmospheres could provide insight into interior processes. If the water outgassing rate can be constrained via water vapor detection on any of the TRAPPIST-1 planets, it will help narrow the allowed space of parameters tested here. If an upper limit on atmospheric water vapor is obtained on e.g. TRAPPIST-1 c, it can be paired with atmospheric chemistry models to calculate the escape rate of water vapor from the potential atmosphere. We can then determine the upper limit on the water outgassing rate by requiring it not exceed the predicted escape rate. This will narrow the allowed parameter space for qualities of TRAPPIST-1 c, such as $Q$ and $m_{\ch{H2O}}^{mantle}$, and may be generalized to the other planets if initial water endowments are assumed to be comparable.

\subsection{Implications for other terrestrial exoplanets}

Our results should apply for a wide variety of terrestrial exoplanets if they have broadly similar conditions to the TRAPPIST-1 planets. In our theoretical model, information specific only to TRAPPIST-1 includes the estimated magma emplacement rates from evolutionary models \citep{Krissansen-Totton.2022}, which depend on planet masses and radii. TRAPPIST-1 specific information includes the XUV flux for the \ch{H2}-escape filter, but this filter had little impact on the allowed parameter space compared to the evolutionary filter, which uses the age of the TRAPPIST-1 system. Therefore, our model results should apply to terrestrial exoplanets with comparable mass, radius, and age to the TRAPPIST-1 planets. More detailed calculations are necessary to determine the extent to which outgassing could support secondary atmospheres against intense atmospheric escape, but our results suggest that present-day outgassing of atmospheric volatiles could play a role on e.g. LHS 3844b \citep{Kreidberg.2019}, GJ 1252b \citep{Crossfield.2022}, Gl 486b \citep{Mansfield.2024}, the LTT 1445 system \citep{Winters.2022}, and the L 98-59 system \citep{Cloutier.2019}. In some cases, tenuous atmospheres are only ruled out because they are not expected to survive intense atmospheric escape rates over extended periods of time \citep{Kreidberg.2019, Mansfield.2024}. Our results suggest that tenuous atmospheres could be supported by present-day outgassing, even in the presence of high escape rates. Thus, present-day tenuous atmospheres should not be ruled out solely on the basis that high escape rates acting over the life of the star would have likely removed an early atmosphere or ocean. 

\section{Conclusions}

We explored a broad phase space of possible present-day water outgassing scenarios on the TRAPPIST-1 planets using observational constraints and theoretical maxima on mantle water content. Our results indicate that if the TRAPPIST-1 planets are terrestrial analogs with a core and mantle, they most likely have (1) low, Mars-like magma emplacement rates, (2) low, Earth-like mantle water mass fractions (although water mass fractions up to 1 wt\% are still plausible), and (3) water outgassing rates $\sim$0.03x Earth's (but rates up to $\sim$8x higher than Earth's are not ruled out). If the TRAPPIST-1 planets have Mars-like magma emplacement rates, their mantle water content is unconstrained and they are likely to have low water outgassing rates ($\sim$0.004x Earth). If they have Io-like magma emplacement rates, their mantle water content is expected to be at or below Earth's and they likely have high water outgassing rates (up to $\sim$7x Earth).

Our plausible ranges of outgassing rates for water and other species can be used to help interpret ongoing observations of the TRAPPIST-1 planets, and our results have implications for future observational and theoretical studies of the TRAPPIST-1 system. Our results indicate high, Io-like volcanism driven by tidal heating is unlikely on the TRAPPIST-1 planets at present day. However, our results include plausible water outgassing rates that could balance the high atmospheric escape rates driven by the parent star, and support water-vapor-containing atmospheres on all of the TRAPPIST-1 planets. These model results, combined with atmospheric observations, could potentially constrain interior characteristics and processes. If a thick water-vapor-containing atmosphere is found on any of the planets, it increases the likelihood of a drier mantle with a high magma emplacement rate; if a tenuous or no water-vapor-containing atmosphere is detected it increases the likelihood of a wetter mantle with a low magma emplacement rate. Future work can further assess potential steady-state water-vapor-containing atmospheres by coupling the outgassing rates provided here to models of photochemistry and atmospheric escape. Our model results are likely applicable to other terrestrial exoplanets with similar mass, radius, and age to the TRAPPIST-1 planets. Ultimately, our results provide a theoretical model that supports the plausibility of habitable zone surface oceans and even tenuous atmospheres on M dwarf terrestrial exoplanets, despite the early super-luminous phase of their host star, and their ongoing high rates of atmospheric escape.

\bigskip

T.B.T. and M.T.G. acknowledge funding from the NSF GRFP (DGE-1762114 and DGE-2140004, respectively). N.F.W. was supported by the NASA Postdoctoral Program. This work is supported by the Virtual Planetary Laboratory, a member of NASA Nexus for Exoplanet System Science (NExSS), funded via the NASA Astrobiology Program grant No. 80NSSC18K0829. We thank Claire Guimond and the other, anonymous reviewer for improving this study.

\bigskip

The Python source code and data in this work are publicly available at Github (\url{https://github.com/trentagon/trout}) and Zenodo (\url{https://doi.org/10.5281/zenodo.15345251}).

\clearpage
\appendix
\section{Additional Posterior Distributions}
\renewcommand{\thefigure}{A\arabic{figure}}
\setcounter{figure}{0}

\begin{figure*}[htbp]
    \centering
    \includegraphics[width=\textwidth]{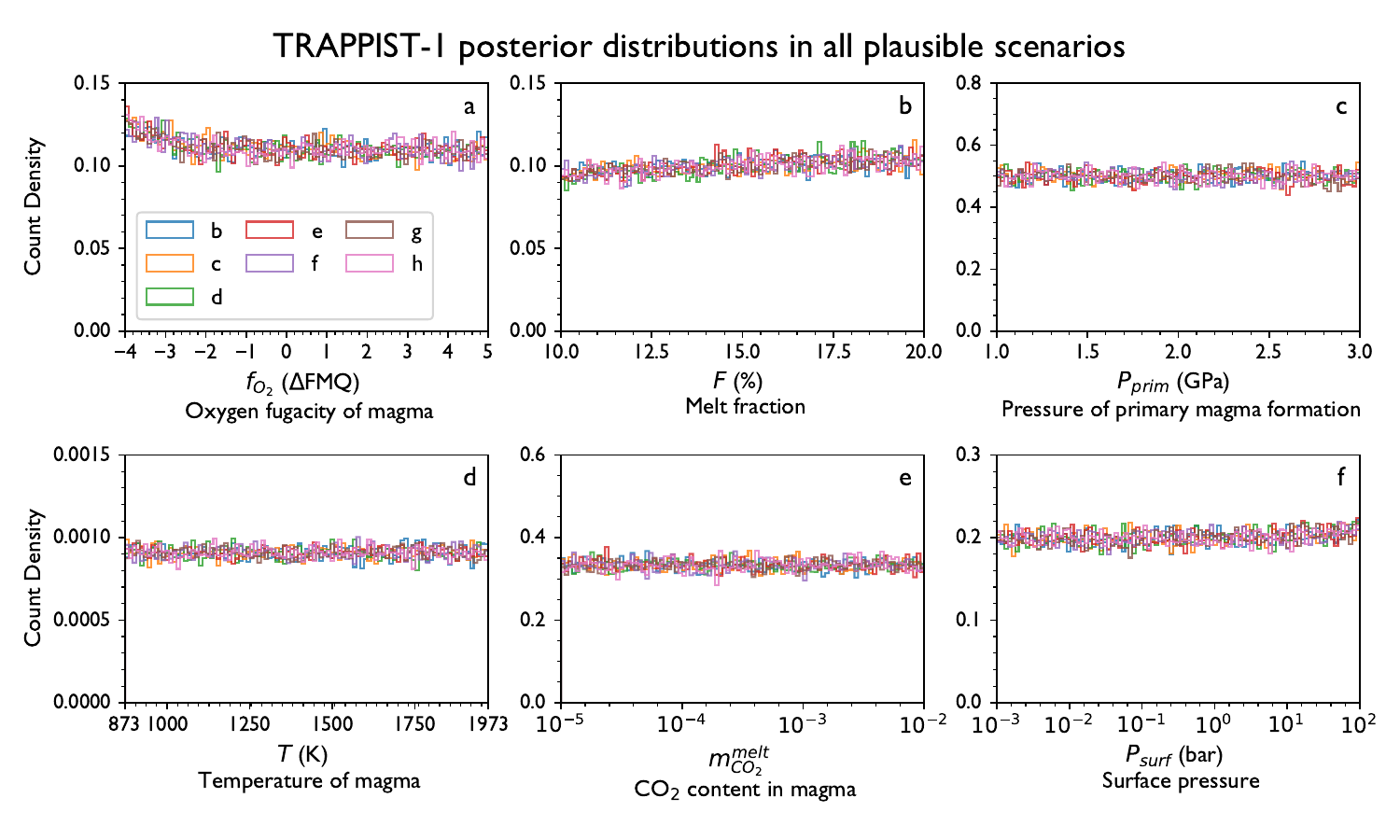}
    \caption{Additional posterior distributions of model values for all TRAPPIST-1 planets in all plausible model scenarios. Plausible model scenarios are derived by first executing the full parameter sweep and then applying empirical and theoretical filters to eliminate model runs, described in Section \ref{sec:filters}.}
    \label{appendix_posteriors}
\end{figure*}

\begin{figure*}[htbp]
    \centering
    \includegraphics[width=\textwidth]{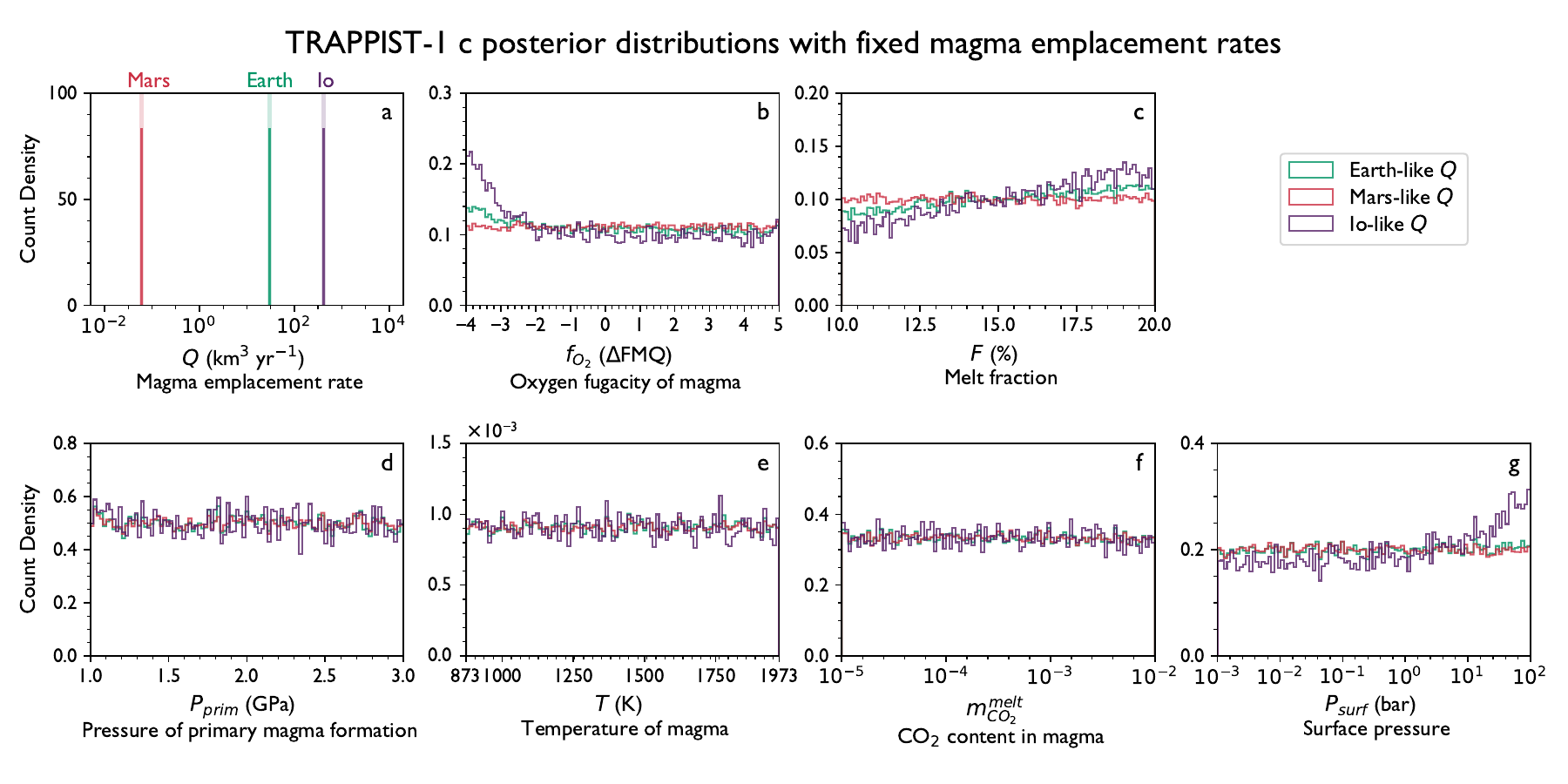}
    \caption{Additional posterior distributions of model values for TRAPPIST-1 c in plausible model scenarios with fixed magma emplacement rates based on those of Earth, Mars, and Io. Plausible model scenarios are derived by first executing the full parameter sweep but with $Q$ fixed to one of the solar system values, then applying empirical and theoretical filters to eliminate model runs that violate known empirical and theoretical constraints, as described in Section \ref{sec:filters}. In this case we apply the filters using the mass and radius of TRAPPIST-1 c as a representative case. (a) Vertical bars are ground truth values of $Q$ from the solar system: Mars, Earth, and Io are 0.05, 31, and 413 km$^3$ yr$^{-1}$, respectively (Section \ref{Q_methods}).}
    \label{appendix_solarcases}
\end{figure*}

Figures \ref{appendix_posteriors} and \ref{appendix_solarcases} show posterior distributions for other model parameters (described in Table \ref{og_params}) that are omitted from the main text because they do not change much from their uniform prior distributions. Figure \ref{appendix_posteriors} shows the additional posterior distributions after Monte Carlo sampling all model parameters 100,000 times, with $Q$ uniformly sampled in base 10 log space within upper limits given in Table \ref{t1_params}. As demonstrated by the flat or near-flat parameter distributions, the filters do not place strong constraints on these model parameters. This indicates that they are less important for determining outgassing rates than the parameters referenced in the main text (magma emplacement rate and mantle water content).

Figure \ref{appendix_solarcases} shows model parameter posterior distributions where the model is Monte Carlo sampled 100,000 times but with $Q$ fixed to values based on solar system observations. These parameter distributions are flat or near-flat with the exceptions of the magma oxygen fugacity (Figure \ref{appendix_solarcases}b) and the magma melt fraction (Figure \ref{appendix_solarcases}c). The magma oxygen fugacity displays a slight preference for low values (i.e. more reduced magma) with Io-like and Earth-like magma emplacement rates. This slight preference is a result of the evolutionary filter, which generally eliminates model runs with high water outgassing rates. With a very reduced magma, the magma water content will be lower (because \ch{H2O} is favored in oxidized magmas), which leads to model runs with lower water outgassing rates that are not eliminated by the evolutionary filter. The melt fraction also displays a slight preference for high values with Io-like and Earth-like magma emplacement rates. This is also a consequence of the evolutionary filter, because high melt fractions result in lower magma water content and thus lower water outgassing rates. Both of these effects are second-order compared to the parameters in the main text and thus only appear as slight deviations from uniform distributions in these specific cases.

\section{Outgassing flux in mol s$^{-1}$ m$^{-2}$}
\renewcommand{\thefigure}{B\arabic{figure}}
\setcounter{figure}{0}

\begin{figure*}[htbp]
    \centering
    \includegraphics[width=0.8\textwidth]{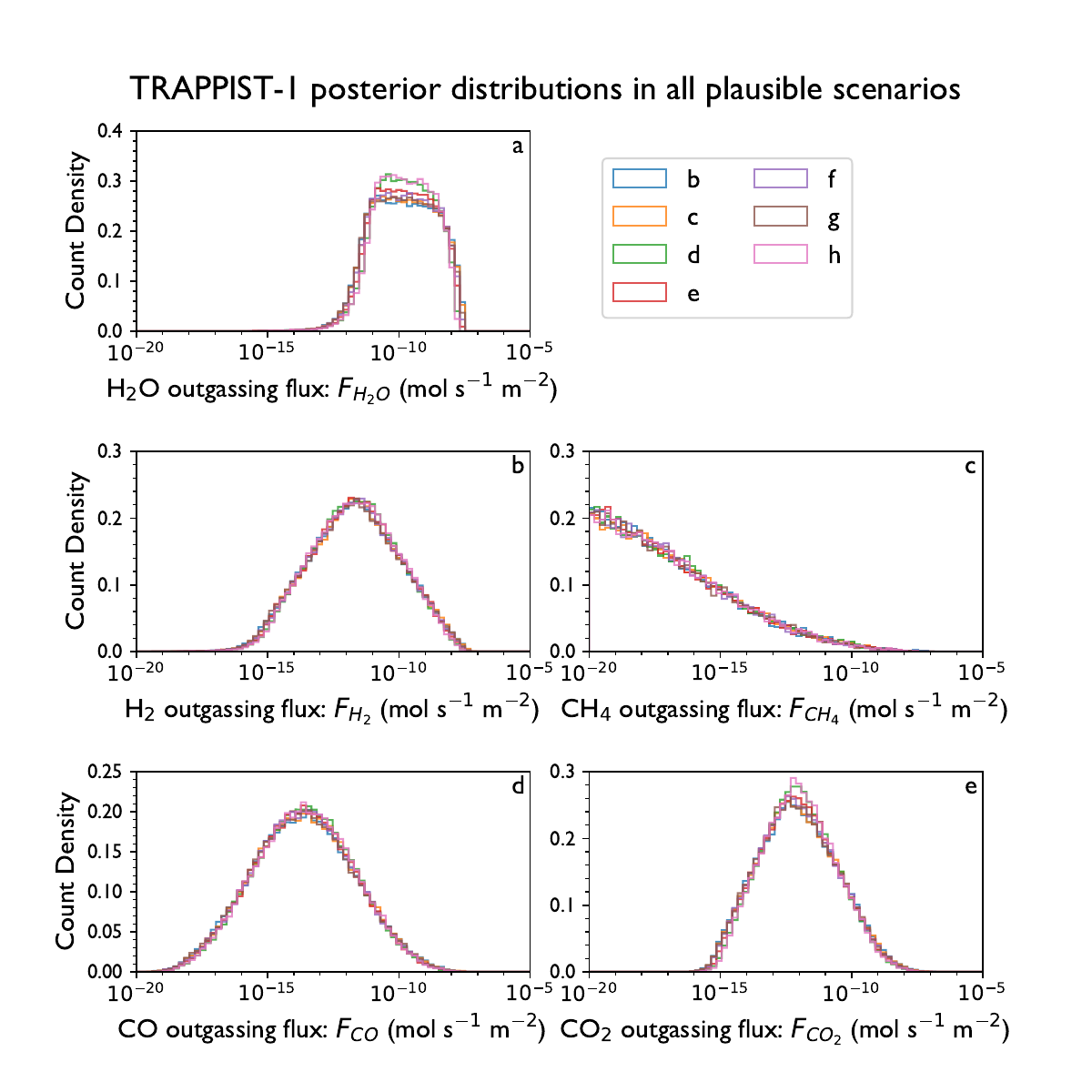}
    \caption{Posterior distributions of outgassing fluxes in mol s$^{-1}$ m$^{-2}$ units for all TRAPPIST-1 planets in all plausible model scenarios. Plausible model scenarios are derived by first executing the full parameter sweep and then applying empirical and theoretical filters to eliminate model runs, described in Section \ref{sec:filters}.}
    \label{appendix_posteriors_fluxunits}
\end{figure*}

\begin{figure*}[htbp]
    \centering
    \includegraphics[width=0.8\textwidth]{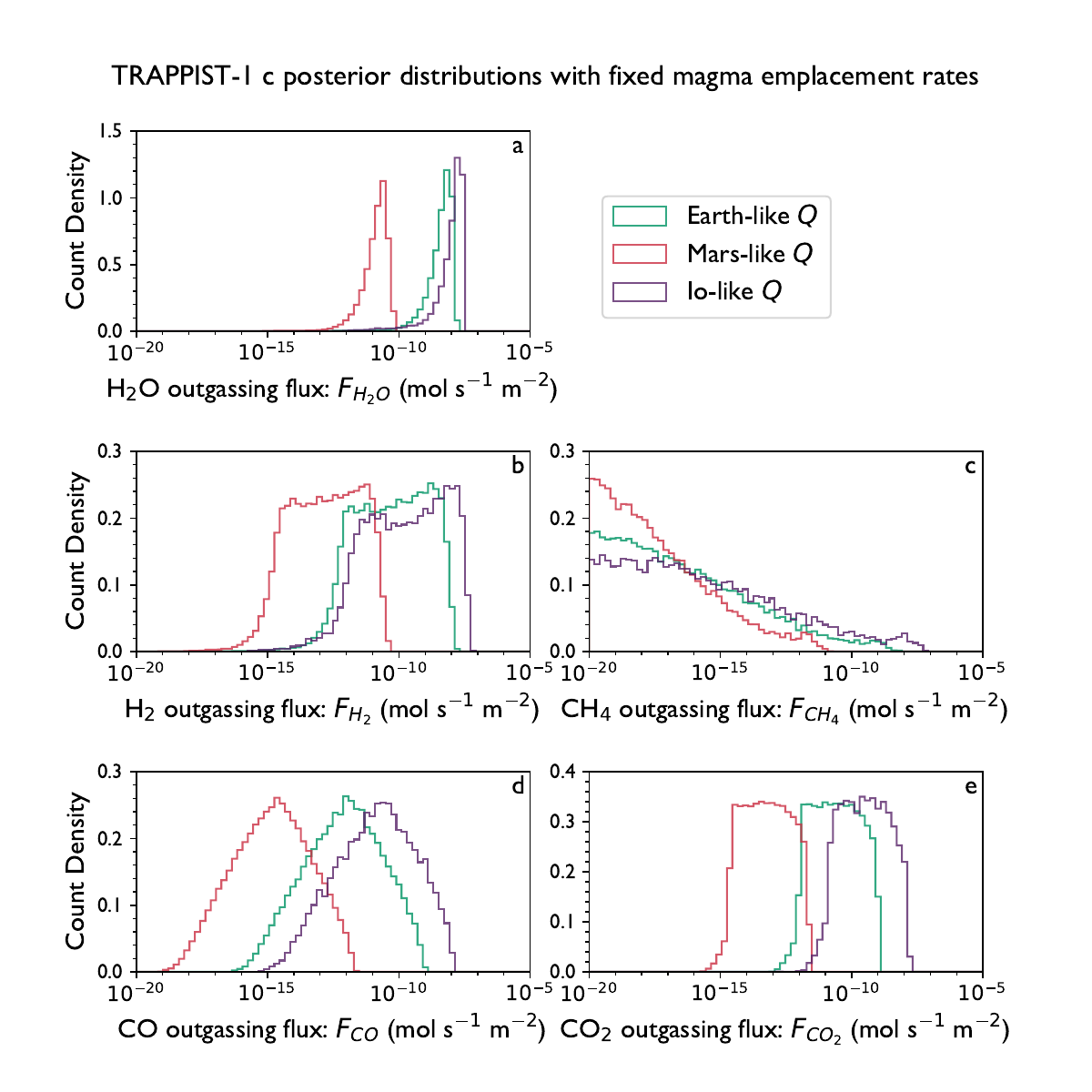}
    \caption{Posterior distributions of outgassing fluxes in mol s$^{-1}$ m$^{-2}$ for TRAPPIST-1 c in plausible model scenarios with fixed magma emplacement rates based on those of Earth, Mars, and Io. Plausible model scenarios are derived by first executing the full parameter sweep but with $Q$ fixed to one of the solar system values, then applying empirical and theoretical filters to eliminate model runs that violate known empirical and theoretical constraints, as described in Section \ref{sec:filters}. In this case we apply the filters using the mass and radius of TRAPPIST-1 c as a representative case. $Q$ for Mars, Earth, and Io is 0.05, 31, and 413 km$^3$ yr$^{-1}$, respectively (Section \ref{Q_methods}).}
    \label{appendix_solarcases_fluxunits}
\end{figure*}

Figures \ref{appendix_posteriors_fluxunits} and \ref{appendix_solarcases_fluxunits} show the posterior distributions for outgassing fluxes in units mol s$^{-1}$ m$^{-2}$, which are provided for compatibility and ease of use in future photochemical modeling studies. The distributions in Figure \ref{appendix_posteriors_fluxunits} correspond directly to those in Figure \ref{t1_posteriors}. The distributions in Figure \ref{appendix_solarcases_fluxunits} correspond directly to those in Figure \ref{t1c_fixedQ_posteriors}. Planet radius values are taken from \citet{Agol.2021}.

\clearpage
\bibliography{troutbib_newcitekey,gto_bibs}{}
\bibliographystyle{aasjournal}

\end{document}